\documentclass[lettersize,journal]{IEEEtran}
\usepackage{amsmath,amsfonts}
\usepackage{algorithmic}
\usepackage{algorithm}
\usepackage{array}
\usepackage[caption=false,font=normalsize,labelfont=sf,textfont=sf]{subfig}
\usepackage{textcomp}
\usepackage{stfloats}
\usepackage{url}
\usepackage{verbatim}
\usepackage{graphicx}
\usepackage{cite}
\usepackage{multirow}
\usepackage{booktabs}
\usepackage{siunitx} 
\usepackage{enumitem} 

\usepackage{siunitx}
\usepackage{eurosym}
\DeclareSIUnit{\eur}{\text{\euro}}

\newcommand\Tstrut{\rule{0pt}{2.1ex}}     

\usepackage{tikz}
\usetikzlibrary{scopes}
\usetikzlibrary{backgrounds}
\usepackage{circuitikz}
\usepackage{rotating}
\usetikzlibrary{arrows,shapes}
\usepackage[normalem]{ulem}
\definecolor{color_WF}{rgb}{0.2,0.1333,0.5333}
\definecolor{color_GFPP}{rgb}{0.800,0.1600,0.0337}

\usepackage{nicefrac}

\definecolor{newtextcolor}{HTML}{bf360c}

\usepackage{xr-hyper}
\makeatletter
\newcommand*{\addFileDependency}[1]{
  \typeout{(#1)}
  \@addtofilelist{#1}
  \IfFileExists{#1}{}{\typeout{No file #1.}}
}
\makeatother
\newcommand*{\myexternaldocument}[1]{%
    \externaldocument{#1}%
    \addFileDependency{#1.tex}%
    \addFileDependency{#1.aux}%
}

\myexternaldocument{main_Online_Companion}

\usepackage{hyperref}

\usepackage{ifthen}
\newcommand{\figspace}{\vspace{-0.275cm}}

\begin{document}

\title{Flexibility of Integrated Power and Gas Systems: Gas Flow Modeling and Solution Choices Matter}

\author{Enrica Raheli, Yannick Werner, and Jalal Kazempour,~\IEEEmembership{Senior Member,~IEEE
} \vspace{-0.4cm} %
\thanks{The first two co-authors contributed equally and are listed alphabetically. All co-authors are with the Department of Wind and Energy Systems at the Technical University of Denmark, 2800 Kgs. Lyngby,
Denmark. Y.\,Werner is also with the Department of Industrial Economics and Technology Management at the Norwegian University of Science and Technology, 7491 Trondheim, Norway.
E-mails: $\{$enrah, yanwe, jalal$\}$@dtu.dk.}}



\vspace{-1cm}
\IEEEaftertitletext{\vspace{-0.6\baselineskip}}
\maketitle

\begin{abstract}
Due to their slow gas flow dynamics, natural gas pipelines function as short-term storage, the so-called \textit{linepack}. By efficiently utilizing linepack, the natural gas system can provide flexibility to the power system through the flexible operation of gas-fired power plants. This requires accurately representing the gas flow physics governed by partial differential equations. Although several modeling and solution choices have been proposed in the literature, their impact on the flexibility provision of gas networks to power systems has not been thoroughly analyzed and compared. This paper bridges this gap by first developing a unified framework. We harmonize existing approaches and demonstrate their derivation from and application to the partial differential equations. Secondly, based on the proposed framework, we numerically analyze the implications of various modeling and solution choices on the flexibility provision from gas networks to power systems. One key conclusion is that relaxation-based approaches allow charging and discharging the linepack at physically infeasible high rates, ultimately overestimating the flexibility.
\end{abstract}

\begin{IEEEkeywords}
Partial differential equations, linepack, discretization, relaxation gap, Weymouth equation 
\end{IEEEkeywords}

\vspace{-0.1cm}

\linespread{0.875}
\section{Introduction}
\label{section:01_Intro}

\IEEEPARstart{P}{ower} and natural gas networks are becoming increasingly coupled due to the extensive use of gas-fired power plants (GFPPs) for electricity generation \cite{Mancarella2016}. Due to their ability to rapidly adjust their power production, these plants are frequently dispatched to counterbalance the fluctuations in variable renewable power generation \cite{Zlotnik_1}. This propagates the variability to the gas system, impacting network management and potentially causing supply shortages or pipeline congestion \cite{Malley2022}. In these cases, power system operators may rely on more expensive measures to replace the GFPPs or even load shedding \cite{Zlotnik_3, Zavala}.

Despite the growing interdependencies, power and gas systems are typically operated independently, with dispatching decisions taken separately due to asynchronous market mechanisms \cite{Byeon2020}. To ensure reliable and cost-efficient operation, several studies have suggested a more coordinated operation of the two systems \cite{Zlotnik_3, Byeon2020, anna2}. The optimal power and gas flow (OPGF) problem should accurately capture the operational constraints and the physics of energy flow in power transmission lines and natural gas pipelines \cite{Roald2020}. 
While steady-state models can represent the fast dynamics of power systems, transient models must be adopted to capture the slow dynamics of gas flows \cite{Malley2022}. 
Due to those slow dynamics, pipelines function similarly to short-term storage. This capability is generally referred to as \textit{linepack} \cite{Liu2011, Menon}. By efficiently utilizing linepack, the gas network can provide flexibility to the power system by enabling rapid changes in the power generation of GFPPs at a low additional gas supply cost. We refer to it as \textit{linepack flexibility}. To harness this flexibility, the efficient coordination of markets and representation of slow gas flow dynamics in the OPGF problem are crucial \cite{Roald2020, Malley2022PMAPS}.

\IEEEpubidadjcol 

While there is a vast literature on the formulation of power flow equations (see \cite{Molzahn2019} for a comprehensive survey), the literature on modeling gas flow equations is comparatively scarce. The flow of natural gas in pipelines is governed by a set of partial differential equations (PDEs), turning the OPGF problem into a PDE-constrained optimization model \cite{lurie2009modeling}. The PDEs can be discretized in time and space dimensions, converting them into nonlinear and nonconvex algebraic equations \cite{THORLEY19873}. We refer to the full set of equations as the so-called  \textit{dynamic} ($\mathrm{DY}$) model. By neglecting the inertia term in the PDEs, it is reduced to the \textit{quasi-dynamic} ($\mathrm{QD}$) model, as it will be discussed later. Further neglecting the linepack flexibility results in the \textit{steady-state} ($\mathrm{ST}$) model.
We refer to the decisions taken on the space and time discretization and the degree of simplification of the PDEs as \textit{modeling choices}.
Irrespective of the modeling choices, the discretized PDEs will constitute nonlinear and nonconvex constraints. Different techniques, including various approximations and relaxations, have been proposed in the literature to solve the resulting OPGF problem. We refer to those techniques as \textit{solution choices}.

Table~\ref{table:review} provides an overview of the existing literature on the optimal operation of integrated power and gas systems, categorized by the modeling and solution choices of the gas flow equations.
\begin{table}[] \label{table:review}
\centering
\caption{Overview of literature on optimal power and gas flow.}
\renewcommand{\arraystretch}{1.3}
\begin{tabular}{lrll}
\hline
\multirow{2}{*}{Model} & \multicolumn{1}{c}{\multirow{2}{*}{Exact}} & \multicolumn{2}{c}{Relaxation} \\ \cline{3-4} 
                  & \multicolumn{1}{c}{}                & Linear                                        & Conic                                     \\ \hline
Dynamic ($\mathrm{DY}$)                &  \cite{Zlotnik_3, CHIANG2016, Mhanna2022a}       &  -                                            &   \cite{Mhanna2022a}                                      \\
Quasi-dynamic ($\mathrm{QD}$)                &   \cite{Mhanna2022}    &  \cite{He2018, ORDOUDIS2019, June2022}        &   \cite{Wang2018, Schwele2019, Chen2019}   \\
Steady-state ($\mathrm{ST}$)                &   \cite{ZENG2016}                   &  \cite{CUI2016183}   &   \cite{YubinHe2018, Singh2019}            \\ \hline
\end{tabular}
\figspace
\end{table}
The table is not exhaustive but provides representative examples of each combination, if available. Both $\mathrm{QD}$ and $\mathrm{ST}$ models have been extensively adopted \cite{Mhanna2022, He2018, ORDOUDIS2019, June2022, Wang2018, Schwele2019, Chen2019, ZENG2016, CUI2016183, YubinHe2018, Singh2019}. They are usually solved either with an outer linear \cite{He2018, ORDOUDIS2019, June2022, CUI2016183} or conic relaxation \cite{ Wang2018, Schwele2019, Chen2019, YubinHe2018, Singh2019}. Both solution choices have been proven to be insufficiently tight under many operating conditions, resulting in physically infeasible solutions \cite{Wang2018, Chen2019, YubinHe2018, SCHWELE2020106776}. Model formulations based on the original nonconvex physics are dominantly solved by interior point methods \cite{Zlotnik_3, CHIANG2016} or successive linear programming \cite{Mhanna2022, Mhanna2022a}.
Except for \cite{Zlotnik_3, Mhanna2022a} and \cite{CHIANG2016}, all the studies in Table~\ref{table:review} use a \qty{1}{\hour} granularity and the original pipeline length for time and space discretization, respectively. Except for \cite{Mhanna2022a}, relaxation approaches have generally been applied to reformulations of the $\mathrm{QD}$ and $\mathrm{ST}$ models based on the so-called Weymouth equation.

Depending on modeling and solution choices, the flexibility provision from the natural gas network to the power system may be over- or under-estimated, causing increased operational costs, physically infeasible schedules, and the curtailment of electrical loads in extreme cases. A systematic comparison of modeling and solution choices for gas flows in integrated power and gas systems concerning flexibility provision is missing in the literature. Different reformulations of PDEs, discretization granularities, and solution techniques make such a comparison challenging. While \cite{ConorThesis} provides the first attempt at offering a comprehensive comparison, the focus has solely remained on the impact of modeling choices on the state variables of gas networks. Furthermore, the impact of solution choices, especially relaxation-based methods, has not been investigated.

This paper bridges those gaps by harmonizing different modeling and solution choices under a unified PDE-based framework, enabling a rigorous comparison. We complement the framework with general recommendations on improving the tightness of existing mathematical formulations and handling numerical issues. Based on that, we analyze the impact of modeling and solution choices on the flexibility provision from power to gas systems using a stylized small-scale case study and a realistic large-scale case study.
The paper provides a comprehensive review of existing gas flow modeling and solution approaches targeting the power systems community analyzing the operation of integrated power and gas systems.

The remainder of this paper is organized as follows. Section~\ref{section:02_modelcomp} introduces the OPGF problem. Section~\ref{section:02_NGmodel} derives and compares the existing gas flow modeling choices. Section~\ref{section:04_NGsol} summarizes different solution choices under a unified PDE-based framework. Section~\ref{sec:computational_recomm} provides general recommendations on modeling gas flows using the PDEs. Section~\ref{section:results} presents an in-depth numerical analysis of the impact of modeling and solution choices on the flexibility provision from gas networks to power systems. Finally, Section~\ref{section:05_conclusion} concludes.

\section{Optimal Power-Gas Flow Problem: Status quo}
\label{section:02_modelcomp}
The multi-period OPGF problem, minimizing the total operating cost of integrated power and gas systems, can be expressed in a generic compact form as
\begingroup
\allowdisplaybreaks
\begin{subequations}\label{eq:OPGF_opt}
\begin{align}
\min_{\mathbf{w},\mathbf{v},\mathbf{x},\mathbf{y}} \quad & f^{\mathbf{P}}(\mathbf{w}) + f^{\mathbf{G}}(\mathbf{x}) \label{eq:OPGF_obj}\\
\textrm{s.t.} \quad & \mathbf{g^{P}(w,v)}\leq \mathbf{0}, \quad \mathbf{g^{G}(x,y)} \leq \mathbf{0}, \label{eq:OPGF_ineq} \\ 
\quad & \mathbf{h^{PB}(w,v)}= \mathbf{0}, \label{eq:OPGF_PB} \\
\quad & \mathbf{h^{PF}(v)} = \mathbf{0}, \label{eq:OPGF_PF} \\
\quad & \mathbf{h^{GB}(x,y,w)}= \mathbf{0}, \label{eq:OPGF_GB} \\
\quad & \mathbf{h^{GF}(y)} = \mathbf{0}, \label{eq:OPGF_GF}
\end{align}
\end{subequations}
where $\mathbf{w}$ and $\mathbf{x}$ represent the vectors of power and gas system decision variables, respectively, including the power output of generators, gas injections at supply nodes, and curtailment of electrical and gas loads. The state variables of power and gas systems are captured by $\mathbf{v}$ and $\mathbf{y}$, respectively, which include voltages, power flows, pressures, and gas flows. The objective function \eqref{eq:OPGF_obj} depends on the power and gas system decision variables $\mathbf{w}$ and $\mathbf{x}$ only.
Constraints~\eqref{eq:OPGF_ineq} represent all technical constraints in power and gas networks, such as generation, voltage, supply, and pressure limits. Constraints~\eqref{eq:OPGF_PB} enforce power balance at each bus in the power grid. Constraints~\eqref{eq:OPGF_PF} represent a variant of the AC power flow constraints. Power and gas systems are linked by the nodal gas balance equations \eqref{eq:OPGF_GB}. Finally, \eqref{eq:OPGF_GF} represents the gas flow in pipelines. While \eqref{eq:OPGF_ineq}, \eqref{eq:OPGF_PB}, and \eqref{eq:OPGF_GB} are convex constraints, \eqref{eq:OPGF_PF} and \eqref{eq:OPGF_GF} are generally nonconvex.

To solve the optimization problem~\eqref{eq:OPGF_opt}, it is a common practice to simplify \eqref{eq:OPGF_PF} and \eqref{eq:OPGF_GF} that govern power and gas flows \cite{ZENG2016, CUI2016183}. In the following sections, we show in detail how the gas flow constraints~\eqref{eq:OPGF_GF} can be formulated, simplified, and solved in various ways proposed in the literature. We refer the interested reader to \cite{Molzahn2019} for a comprehensive summary of power flow modeling.
\section{Gas Flow: Modeling Choices}
\label{section:02_NGmodel}
Natural gas is primarily transported over long distances through pressurized pipelines. The one-dimensional dynamic flow of natural gas along the pipeline axis is governed by a set of PDEs known as Euler's equations. For high-pressure natural gas transmission pipelines, a set of simplifying assumptions is generally adopted, which allows rewriting the original PDEs \cite{ConorThesis} as
\begingroup
\allowdisplaybreaks
\begin{align}
    & \underbrace{\frac{\partial \pi}{\partial t}}_{\text{Linepack change}} + \underbrace{\frac{c^2}{A}\frac{\partial m}{\partial x}}_{\text{Net mass outflow}} = 0, \label{eq:mass} \\
    & \underbrace{\frac{\partial m}{\partial t}}_\text{Inertia term} + \underbrace{A \frac{\partial \pi}{\partial x}}_\text{Pressure gradient} + \underbrace{\frac{\lambda c^2}{2DA} \frac{m |m|} {\pi}}_\text{Friction force} = 0. \label{eq:momentum}
\end{align}
Equations \eqref{eq:mass}--\eqref{eq:momentum} represent the conservation of mass and momentum equations, respectively. The symbols $t$ and $x$ denote the time and space dimensions, respectively. The symbols $\pi$ and $m$ are the pressure and gas mass flow rates. Constants $D$, $A$, and $\lambda$ are pipeline-specific and denote the diameter, cross-sectional area, and friction factor. Finally, constant $c$ is the speed of sound in the transported gas. We refer the interested reader to Section~$\mathrm{OC.1}$ of the online companion \cite{Online_Companion} for a detailed derivation of those PDEs. The meaning of each term indicated by the curly brackets is explained throughout this section. 

Equations \eqref{eq:mass}--\eqref{eq:momentum} are the basis of the $\mathrm{DY}$ model, which will be explained in Section \ref{subsection:02_dynamic}. 
Based on simplifying assumptions on the contribution of the inertia term and friction force on the pressure gradient in~\eqref{eq:momentum}, the $\mathrm{DY}$ model can be transformed into the $\mathrm{QD}$ model, which will be discussed in Section~\ref{subsection:02_quasidynamic}. By assuming that the intertemporal change in linepack in \eqref{eq:mass} is zero, the $\mathrm{QD}$ model can be further simplified to the $\mathrm{ST}$ model, which will be described in Section~\ref{subsection:02_steadystate}. Before elaborating on the individual models and their impact on the gas network flexibility, we show how to transform the PDEs into tractable algebraic expressions.

\subsection{Discretization of the PDEs: The dynamic ($\mathrm{DY}$) model}\label{subsection:02_dynamic}
Since there is no known analytical solution to the presented system of PDEs, numerical methods must be used to approximate and solve \eqref{eq:mass}--\eqref{eq:momentum}. These methods generally transform the PDEs into algebraic expressions with discrete time and space dimensions, e.g., using finite difference methods that approximate the partial derivatives \cite{ConorThesis, kiuchi1994implicit}. In the following, we describe how the implicit cell-centered method proposed in \cite{kiuchi1994implicit} can be used to discretize the PDEs \eqref{eq:mass}--\eqref{eq:momentum}. 

The spatial derivatives are discretized by dividing the original pipelines into segments of equal length $\Delta x$ by introducing auxiliary nodes. If an original pipeline is shorter than the chosen space discretization $\Delta x$, we do not discretize it and use its original length. Let $i,j \in \mathcal{I}$ denote the nodes of the gas network (including auxiliary nodes) and $(i,j) \in \mathcal{P}$ the discretized pipeline segments connecting them. Then, we denote the length of pipeline segment $(i,j)$ by $\Delta x_{ij}$. Hereafter, we will not refer specifically to pipeline segments and auxiliary nodes but simply to pipelines and nodes.
The temporal derivatives are discretized using a uniform segment length $\Delta t$. Based on the implicit cell-centered method, the conservation of mass and momentum equations~\eqref{eq:mass}--\eqref{eq:momentum} for each pipeline $(i,j)$ and time step $t$ are discretized as
\begingroup
\allowdisplaybreaks
\begin{align}
    & \underbrace{\frac{\pi_{ij,t}-\pi_{ij,t-1}}{\Delta t}}_\text{$\mathrm{DY}$, $\mathrm{QD}$} + \underbrace{\frac{c^2}{A_{ij}}\frac{m_{ij,t}^{\mathrm{out}}-m_{ij,t}^{\mathrm{in}}}{\Delta x_{ij}}}_\text{$\mathrm{DY}$, $\mathrm{QD}$, $\mathrm{ST}$} = 0, \label{eq:mass_d}\\
    & \underbrace{\frac{m_{ij,t}-m_{ij,t-1}}{\Delta t}}_\text{$\mathrm{DY}$} + \underbrace{A_{ij} \frac{\pi_{j,t}-\pi_{i,t}}{\Delta x_{ij}}}_\text{$\mathrm{DY}$, $\mathrm{QD}$, $\mathrm{ST}$} +  \underbrace{\frac{\lambda_{ij} c^2}{2D_{ij}A_{ij}}\frac{m_{ij,t}|m_{ij,t}|}{\pi_{ij,t}}}_\text{$\mathrm{DY}$, $\mathrm{QD}$, $\mathrm{ST}$}= 0. \label{eq:momentum_d}
\end{align}

The average pressure\footnote{The average pressure in the pipeline can be expressed more accurately by a nonlinear expression (see \cite{SAEDI2021117598, ZHANG20222220}). To avoid adding additional nonlinearities, the arithmetic average is used in this paper. However, if the $\mathbf{P}$-$\mathrm{NLP}$ solution approach described in Section \ref{sec:NLP} is used, the nonlinear average pressure expression may be adopted.} $\pi_{ij,t}$ and mass flow $m_{ij,t}$ are defined as
\begingroup
\allowdisplaybreaks
\begin{subequations} \label{eq:m1}
\begin{align}
    \pi_{ij,t} &= \frac{\pi_{i,t}+\pi_{j,t}}{2}, \label{eq:pr_ave}\\
    m_{ij,t} &= \frac{m_{ij,t}^{\mathrm{in}}+m_{ij,t}^{\mathrm{out}}}{2}, \label{eq:m_ave}
\end{align}
\end{subequations}
where $m_{ij,t}^{\mathrm{in}}$ and $m_{ij,t}^{\mathrm{out}}$ are the mass flow at the start and the end of the pipeline, respectively.

The set of algebraic equations \eqref{eq:mass_d}--\eqref{eq:m1} approximates the original PDEs \eqref{eq:mass}--\eqref{eq:momentum} with increasing accuracy as temporal and spatial discretization are refined. The impact of discretization granularity on the optimal gas flows and pressures is analyzed in \cite{ConorThesis}. For completeness, we recap the main findings in Section~$\mathrm{OC.2}$ of the online companion \cite{Online_Companion}, e.g., that a discretization in time has a greater impact on the approximation accuracy of the PDEs than a discretization in space.\footnote{The interested reader is referred to \cite{8768025, Steinle} to learn about more general approaches to choose the discretization based on the pipeline characteristics.}

The curly brackets below equations~\eqref{eq:mass_d}--\eqref{eq:momentum_d} indicate what terms appear in which of $\mathrm{DY}$, $\mathrm{QD}$, and $\mathrm{ST}$ models. The $\mathrm{DY}$ model presents the full set of PDEs~\eqref{eq:mass_d}--\eqref{eq:m1}. In the following two sections, we elaborate on the assumptions taken to simplify the $\mathrm{DY}$ model to the $\mathrm{QD}$ and $\mathrm{ST}$ models.

\subsection{The quasi-dynamic ($\mathrm{QD}$) model} \label{subsection:02_quasidynamic}
It has been shown, e.g., in \cite{Osiadacz1984} and \cite{Herran2009}, that the contribution of the inertia term to the pressure drop in the conservation of momentum equation~\eqref{eq:momentum} is relatively low compared to the friction force (less than \qty{1}{\percent}). In \cite{Hennings2021}, the magnitude of the inertia term in real-world situations is assessed using a large set of historical data for the German gas network with a high temporal resolution of $3$ minutes.
The contribution of the inertia term becomes relevant ($>\qty{0.5}{\bar}$) only when very fast dynamics occur (e.g., sudden shut-down or start-up of a large gas-fired generator), which happens very rarely \cite{Hennings2021}. As shown in Section~$\mathrm{OC.2}$ of the online companion \cite{Online_Companion}, one can verify their empirical findings with the solutions of the optimal gas flow problem. Based on those observations, the $\mathrm{DY}$ model can be simplified by neglecting the inertia term in \eqref{eq:momentum_d}. This reads as
\begin{align}
    & A_{ij} \frac{\pi_{j,t}-\pi_{i,t}}{\Delta x_{ij}} +  \frac{\lambda_{ij} c^2}{2D_{ij}A_{ij}}\frac{m_{ij,t}|m_{ij,t}|}{\pi_{ij,t}}= 0. \label{eq:momentum_d_QD}
\end{align}

This results in the $\mathrm{QD}$ model, which expresses the gas flow dynamics for each pipeline by \eqref{eq:mass_d}, \eqref{eq:m1}, and \eqref{eq:momentum_d_QD}.

\subsection{The steady-state ($\mathrm{ST}$) model} \label{subsection:02_steadystate}
For further simplifying the PDEs, it is often assumed that the network operates in steady-state conditions, setting all temporal derivatives in \eqref{eq:mass}--\eqref{eq:momentum} to zero. The conservation of mass equation~\eqref{eq:mass} reduces to stating that, for each pipeline, the gas inflow is equal to the outflow, i.e., 
\begin{align}
    m_{ij,t}^{\mathrm{in}} = m_{ij,t}^{\mathrm{out}} = m_{ij,t}, \label{eq:mass_d_ST}
\end{align}
meaning that it is constant along the pipeline. The $\mathrm{ST}$ model is then given by~\eqref{eq:pr_ave}, \eqref{eq:momentum_d_QD}, and \eqref{eq:mass_d_ST}.

Removing the temporal derivatives uncouples the PDEs to be considered independent for each time step. Consequently, time discretization does not affect the PDEs in the $\mathrm{ST}$ model. Similarly, as the flow is assumed to be constant along the pipeline, space discretization does not affect the PDEs either.

\subsection{Linepack modeling and Weymouth equation} \label{subsection:linepack}
For the $\mathrm{DY}$, $\mathrm{QD}$ and $\mathrm{ST}$ models, the linepack $h_{ij,t}$, i.e., the amount of gas in pipeline $(i,j)$ at time step $t$, can be approximated as
\begin{align}
    & h_{ij,t} = \frac{A_{ij} \Delta x_{ij}}{c^2}\pi_{ij,t},\label{eq:lp}
\end{align}
which is  proportional to the average pressure $\pi_{ij,t}$. Note that \eqref{eq:lp} approximates the linepack with increasing accuracy as the spatial discretization is refined.
For the PDE-based version of the $\mathrm{DY}$, $\mathrm{QD}$, and $\mathrm{ST}$ models presented in Sections~\ref{subsection:02_dynamic}, \ref{subsection:02_quasidynamic}, and  \ref{subsection:02_steadystate}, the linepack is not an optimization variable, but can be retrieved ex-post from the optimal average pressure $\pi_{ij,t}$.
Note that from a mathematical point of view, the linepack can also be calculated ex-post for the $\mathrm{ST}$ model. However, since the first term of \eqref{eq:mass_d} is neglected, the impact of the intertemporal change in linepack on the pipeline inflow and outflow is not captured (see \eqref{eq:mass_d_ST}). 

For the $\mathrm{DY}$ and $\mathrm{QD}$ models, the flexibility from a change in linepack can be explicitly expressed by substituting \eqref{eq:lp} into \eqref{eq:mass_d}, resulting in a storage-like constraint as
\begin{align}
    & h_{ij,t}= h_{ij,t-1}+\Delta t (m_{ij,t}^{\mathrm{in}} - m_{ij,t}^{\mathrm{out}}). \label{eq:lp_st}
\end{align}

An equivalent reformulation of the $\mathrm{QD}$ model, often used in the literature in the context of integrated power and gas systems analysis \cite{Mhanna2022, He2018, ORDOUDIS2019, June2022, Wang2018, Schwele2019, Chen2019}, is obtained by explicitly including the linepack as a variable and incorporating  \eqref{eq:lp} into the optimization problem. In these studies, the momentum equation~\eqref{eq:momentum_d_QD} is usually reformulated into the Weymouth equation\footnote{In this paper, we will always use the term ``Weymouth equation'' to indicate the general flow equation of gas in pipelines. While this terminology originates from the definition of the friction factor after Weymouth, it has also been used in the power systems community when other definitions have been deployed. To avoid confusion with the original PDEs governing the gas flow, we keep the term Weymouth equation to refer to Equation~\eqref{eq:weymouth}.}, which is obtained by plugging \eqref{eq:pr_ave} into~\eqref{eq:momentum_d_QD}, rearranging to
\begin{align}
    & m_{ij,t}|m_{ij,t}|= \frac{D_{ij} A^2_{ij}}{\lambda_{ij} c^2 \Delta x_{ij}} (\pi_{i,t}^2-\pi_{j,t}^2). \label{eq:weymouth}
\end{align}

The $\mathrm{QD}$ model can therefore be equivalently expressed as \eqref{eq:m1} and \eqref{eq:lp}--\eqref{eq:weymouth}. A similar reformulation can be obtained for the $\mathrm{ST}$ model using \eqref{eq:mass_d_ST} and \eqref{eq:weymouth}, i.e., the Weymouth equation with a constant flow along the pipeline. This formulation of the $\mathrm{ST}$ model has been most commonly adopted in the literature \cite{ZENG2016, CUI2016183, YubinHe2018, Singh2019}.

While we find the direct use of the discretized PDEs~\eqref{eq:mass_d}--\eqref{eq:momentum_d} in the optimization models beneficial from a mathematical and computational point of view, the presented reformulations, including a linepack variable and the Weymouth equation, are often useful for interpretations. This is mainly due to \eqref{eq:lp}--\eqref{eq:lp_st}, which enable storage-like interpretations of the gas inside a pipeline. Hence, due to the proportional relationship between average pressure and linepack in~\eqref{eq:lp}, changes in average pressure are also referred to as changes in linepack.
\section{Gas Flow: Solution Choices}
\label{section:04_NGsol}
This section elaborates on different solution choices for the optimization problem \eqref{eq:OPGF_opt}, which are independent of the modeling approach, i.e., $\mathrm{DY}, \mathrm{QD}$, and $\mathrm{ST}$, and discretization granularity, i.e., $\Delta t$ and $\Delta x$. Solving \eqref{eq:OPGF_opt} by replacing~\eqref{eq:OPGF_GF} with \eqref{eq:mass_d}--\eqref{eq:m1} is nontrivial as \eqref{eq:momentum_d} is nonlinear and nonconvex due to the term related to the friction force.
Following \cite{ConorThesis}, we start by replacing this nonlinear and nonconvex term with the auxiliary variable $\gamma_{ij,t}$, such that 
\begin{equation}\label{eq:gamma_def}
    \gamma_{ij,t} = \frac{m_{ij,t}|m_{ij,t}|}{\pi_{ij,t}},
\end{equation}
turning the conservation of momentum~\eqref{eq:momentum_d} into a linear constraint. Afterwards, \eqref{eq:gamma_def} remains the only nonlinear and nonconvex constraint in~\eqref{eq:OPGF_GF}. For notational brevity, we drop subscript $ij,t$.

Depending on the solution choice $\mathbf{S}$, which we will introduce in the following, the right-hand-side of \eqref{eq:gamma_def} is replaced by a possibly convex feasible set $\mathcal{F}^{\mathbf{S}}$, i.e., 
\begin{align}
    & \gamma \in \mathcal{F}^{\mathbf{S}}. \label{eq:gamma_def_set}
\end{align}


We provide a full description of the OPGF problem~\eqref{eq:OPGF_opt}, which we use for this work, in Appendix~\ref{section:appendix_OPT}. As our main focus is gas flow modeling, we adopt a DC power flow approximation to simplify~\eqref{eq:OPGF_PF}. However, all the modeling and solution choices presented in this paper are generally compatible with all AC power flow constraint variations.

Based on the generic feasible set~\eqref{eq:gamma_def_set}, we define the solution choice dependent OPGF problem $\mathbf{P}$-$\mathbf{S}$, where $\mathbf{P} \in \{ \mathrm{DY}, \mathrm{QD}, \mathrm{ST} \}$:
\begin{equation}
    \operatorname{\mathbf{P}-\mathbf{S}}: \eqref{eq:ob_OPGF}-\eqref{eq:OPGF_PDEs}, \eqref{eq:gamma_def_set}, \label{eq:problem_p_s}
\end{equation}
where the mathematical model \eqref{eq:ob_OPGF}--\eqref{eq:OPGF_PDEs} is presented in Appendix~\ref{section:appendix_OPT} and \eqref{eq:gamma_def_set} is replaced by the respective feasible set representing the solution choice $\mathbf{S}$.

In the following, we will show how the most prevalent solution approaches in the literature can be expressed by different formulations of \eqref{eq:gamma_def_set}. We restrict ourselves to providing short descriptions here while giving further details in Section~$\mathrm{OC.3}$ of the online companion \cite{Online_Companion}.

\subsection{Nonlinear programming}
\label{sec:NLP}
A natural idea is to directly use \eqref{eq:gamma_def} without any reformulation to replace~\eqref{eq:gamma_def_set}, equivalent to the original formulation. We call this model $\mathbf{P}$-$\mathrm{NLP}$. It can be solved to local optimality using interior point methods. Some recent off-the-shelf solvers, e.g., the open-source solver Ipopt \cite{Waechter2005}, are directly able to handle such terms and have been successfully applied to integrated power and gas systems studies, e.g., in \cite{Zlotnik_3} and \cite{CHIANG2016}. 

\subsection{Sequential linear programming}
\label{sec:SLP}
Like interior point methods, sequential linear programming can be applied to find a locally optimal solution to problem~\eqref{eq:problem_p_s}, e.g., as in \cite{Mhanna2022}, \cite{ ConorThesis}, and \cite{LOHR2020106724}. In this case, a series of linear problems are solved that utilize the first-order Taylor series approximation of the nonlinear and nonconvex constraint~\eqref{eq:gamma_def} around the solution of the last iteration:
\begin{align}
\gamma^{k+1} = & \frac{m^k|m^k|}{\pi^k} + \frac{2|m^k|}{\pi^k}(m - m^k) - \frac{m^k|m^k|}{(\pi^k)^2}(\pi - \pi^k) \nonumber \\
= & \frac{2|m^k|}{\pi^k} m - \frac{m^k|m^k|}{(\pi^k)^2}\pi \label{eq:gamma_def_SCP},
\end{align}
where $m^k$ and $\pi^k$ are the optimal mass flow and pressure values at iteration $k$. 
We refer to this algorithm as $\mathbf{P}$-$\mathrm{SLP}$. Appendix~\ref{sec:SLP_app} provides further information on the algorithm.

\subsection{Nonconvex quadratic programming}
Another approach that has rarely been used is to replace the absolute value in~\eqref{eq:gamma_def} by introducing binary variable $z$ and nonnegative variables $\gamma^+, \gamma^-$, $m^+$, and $m^-$ to separate both flow directions\footnote{Here, we define the positive flow direction equal to the direction of the pipeline, e.g., when a pipeline goes from node 1 to 2.}:
\begin{subequations}\label{eq:gamma_def_MIBLP}
\begin{align}
& 0 \leq \gamma^+ \leq z \overline{\Gamma}, & \quad & 0 \leq \gamma^- \leq (z-1) \underline{\Gamma}, \label{eq:gamma_bounds_MIBLP} \\
& 0 \leq m^+ \leq z \overline{M}, & \quad & 0 \leq m^- \leq (z-1) \underline{M}, \label{eq:m_bounds_MIBLP} \\
& \gamma = \gamma^+ - \gamma^-, & \quad & m = m^+ - m^-, \label{eq:gamma_def_twoway_MIBLP}\\
& \gamma^+ = \frac{m^+m^+}{\pi}, & \quad & \gamma^- = \frac{m^-m^-}{\pi}, \label{eq:gamma_def_flow_MIBLP}
\end{align}
\end{subequations}
where $\overline{M} \geq 0$ and $\underline{M} \leq 0$ denote the maximum flow in each direction, respectively. Similarly, $\overline{\Gamma} \geq 0$ and $\underline{\Gamma} \leq 0$ describe the maximum value of the nonnegative auxiliary variables $\gamma^+$ and $\gamma^-$, respectively. In Section~\ref{sec:bounds_mass_flow}, we will elaborate on how tight bounds can be derived. The resulting nonconvex quadratic program is termed $\mathbf{P}$-$\mathrm{NQP}$. This problem can be hypothetically solved to global optimality by the nonconvex feature of Gurobi \cite{gurobi}. In practice, however, we cannot solve it even for a small case study. Therefore, we do not consider this solution choice further.

\subsection{Mixed-integer conic relaxation}  
\label{sec:MISOCP}
A conic relaxation of the Weymouth equation~\eqref{eq:weymouth} has been frequently adopted in the literature \cite{Wang2018, Schwele2019, Chen2019, YubinHe2018, Singh2019}. Similarly, it is possible to relax equality constraints~\eqref{eq:gamma_def_flow_MIBLP} as
\begin{subequations}\label{eq:gamma_def_MISOCP}
\begin{align}
    & \eqref{eq:gamma_bounds_MIBLP}-\eqref{eq:gamma_def_twoway_MIBLP}, & \quad & \label{eq:gamma_def_twoway_MISOCP}\\
    & \gamma^+ \geq \frac{m^+m^+}{\pi}, & \quad & \gamma^- \geq \frac{m^-m^-}{\pi}. \label{eq:gamma_def_flow_MISOCP}
\end{align}
\end{subequations}

As $\gamma^+, \gamma^- \geq 0$ and $\pi > 0$, \eqref{eq:gamma_def_flow_MISOCP} can be reformulated into (rotated) second-order cone constraints.  We refer to the resulting problem as $\mathbf{P}$-$\mathrm{MISOCP}$. In contrast to the commonly applied second-order cone relaxation based on the Weymouth equation~\eqref{eq:weymouth}, this formulation does not require additional auxiliary variables and convex relaxations for the reformulation of the difference in squared pressures \cite{Schwele2019} and therefore exhibits a tighter formulation.\footnote{We note that this is not required for the $\mathrm{ST}$ model as the squared pressures can be substituted by their linear counterparts. This is possible because the pressure variables solely appear in the definition of the average pressure \eqref{eq:pr_ave} and linepack~\eqref{eq:lp}, which are not part of the $\mathrm{ST}$ model (see Section~\ref{subsection:02_steadystate})} It further allows application to the $\mathrm{DY}$ model, which cannot be expressed using the Weymouth equation~\eqref{eq:weymouth}.

Constraints~\eqref{eq:gamma_def_flow_MISOCP} will generally not be binding at the optimal solution \cite{Wang2018, Chen2019, YubinHe2018, SCHWELE2020106776}.
Consequently, the solution violates the gas flow physics governed by the PDEs. We will discuss the consequences of an inexact relaxation of the gas flow physics in Section~\ref{sec:relax_gap_quantification}.

Further attention has been dedicated to using the concave relaxation of the Weymouth equation~\eqref{eq:weymouth} in addition to the conic one. To solve the resulting nonconvex problem, \cite{Wang2018} and \cite{YubinHe2018}  use a convex-concave procedure based on sequential linear programming, while McCormick envelopes and bound tightening have been applied in \cite{Chen2019}. These methods reduce the relaxation gap to zero but at the expense of a globally optimal solution. Additionally, these methods tend to be computationally expensive. Therefore, we do not see an advantage over using interior point or sequential linear programming methods from the start. 

We propose to strengthen the relaxation~\eqref{eq:gamma_def_flow_MISOCP} by introducing a linear overestimator, which substantially reduces the feasible region of problem $\mathbf{P}$-$\mathrm{MISOCP}$:
\begin{align}
    & \gamma^+ \leq m^+ \frac{\overline{M}}{\widehat{\Pi}^{+}},
    & \gamma^- \leq m^- \frac{\underline{M}}{\widehat{\Pi}^{-}}, \label{eq:LO_def}
\end{align}
where $\widehat{\Pi}^{+}$ and $\widehat{\Pi}^{-}$ are the maximum feasible pipeline pressure differences in positive and negative flow direction, respectively. 
We consider the inclusion of the linear overestimator as default while explicitly denoting its neglect. To our knowledge, neither the presented conic relaxation nor the linear overestimator has been applied in the literature.

\subsection{Mixed-integer linear relaxation}
\label{sec:MILP}
Another approach that has been widely adopted in the literature \cite{He2018, ORDOUDIS2019, June2022, CUI2016183, Tomasgard2007} is to use an outer approximation of the Weymouth equation~\eqref{eq:weymouth} based on the first-order Taylor series expansion.
The same approach can be applied to~\eqref{eq:gamma_def_flow_MIBLP} (cf. Section~\ref{sec:SLP}) using a set of carefully chosen linearization points $(\widetilde{m},\widetilde{\pi}) \in \mathcal{U}^{\nicefrac{+}{-}}$:
\begin{subequations}\label{eq:gamma_def_MILP}
\begin{align}
    & \eqref{eq:gamma_bounds_MIBLP}-\eqref{eq:gamma_def_twoway_MIBLP}, & \quad & \label{eq:gamma_def_twoway_MILP} \\
    & \gamma^+ \geq \frac{2|\widetilde{m}|}{\widetilde{\pi}} m^+ - \frac{\widetilde{m}|\widetilde{m}|}{(\widetilde{\pi})^2}\pi, & \forall & (\widetilde{m}, \widetilde{\pi}) \in \mathcal{U}^+, \label{gamma_def_pos_MILP} \\ 
    & \gamma^- \geq \frac{2|\widetilde{m}|}{\widetilde{\pi}} m^- - \frac{\widetilde{m}|\widetilde{m}|}{(\widetilde{\pi})^2}\pi, & \forall & (\widetilde{m}, \widetilde{\pi}) \in \mathcal{U}^-. \label{gamma_def_neg_MILP}
\end{align}
\end{subequations}

The resulting model, $\mathbf{P}$-$\mathrm{MILP}$, is a mixed-integer linear program. It constitutes an outer approximation to the conic relaxation~\eqref{eq:gamma_def_flow_MISOCP}.\footnote{This is not the case when applying the outer linear and conic relaxations to the Weymouth equation~\eqref{eq:weymouth}. In that case, due to the additionally required convex relaxation of the squared pressures (see Section~\ref{sec:MISOCP}), the outer linear relaxation is usually tighter.} 
The feasible region can be again tightened by including the linear overestimator~\eqref{eq:LO_def}, which we consider as the default if not denoted otherwise.

\subsection{Polyhedral envelopes}
\label{sec:PELP}
It is proposed in~\cite{Mhanna2022} to derive a polyhedral envelope for the Weymouth equation~\eqref{eq:weymouth} based on the intersection of halfspaces defined by the first-order Taylor series expansion.
Similarly, a polyhedral envelope to~\eqref{eq:gamma_def} can be defined as
\begin{subequations}\label{eq:gamma_PELP}
\begin{align} 
    & \gamma \geq \frac{2|\widetilde{m}|}{\widetilde{\pi}} m - \frac{\widetilde{m}|\widetilde{m}|}{(\widetilde{\pi})^2}\pi, & \forall & (\widetilde{m}, \widetilde{\pi}) \in \mathcal{U}, \\
    & \gamma \leq \frac{2|\widetilde{m}|}{\widetilde{\pi}} m - \frac{\widetilde{m}|\widetilde{m}|}{(\widetilde{\pi})^2}\pi, & \forall & (\widetilde{m}, \widetilde{\pi}) \in \mathcal{O},
\end{align}
\end{subequations}
where $\mathcal{U}$ and $\mathcal{O}$ are sets of carefully chosen linearization points. We refer to the resulting linear problem as $\mathbf{P}$-$\mathrm{PELP}$, which is computationally very efficiently solvable to global optimality. However, similar to other relaxation-based models, it often yields a solution at which~\eqref{eq:gamma_def} is not satisfied, i.e., it violates the gas flow physics.

\subsection{Piecewise linear approximation}
\label{sec:PWL_app}
It has been frequently proposed in the literature to adopt a one-dimensional piecewise linearization of both sides of the Weymouth equation~\eqref{eq:weymouth}, see, e.g., \cite{Correa-Posada2015, He_ADMM_2017}. Similarly, although more complex, a two-dimensional piecewise linearization of~\eqref{eq:gamma_def} can be derived. Here, we adopt the logarithmic disaggregated convex combination model presented in \cite{Vielma2009}, as it is computationally preferable to other piecewise linearization methods \cite{Geissler2011}. We term the resulting model $\mathbf{P}$-$\mathrm{PWL}$. The solution to this model will generally inhibit an approximation gap, which, similar to the relaxation-based models, implies a violation of the gas flow physics. 
However, the magnitude of the error is usually smaller and controllable by the number of linearization points.
Even with the minimum number of linearization points, we could not solve the resulting problem for a small case study. Therefore, we do not consider this modeling approach further and refer the interesting reader to \cite{Geissler2011} to review existing approaches, which have also been applied to gas flow modeling.
\section{General recommendations}
\label{sec:computational_recomm}
Solving the OPGF problem is numerically challenging due to the PDEs governing the gas flow. In Section~$\mathrm{OC.4}$ of the online companion \cite{Online_Companion}, we propose a per-unit conversion of the OPGF problem, which we find very effective in reducing computational complexity.
As our two additional recommendations, Section~\ref{sec:bounds_mass_flow} derives tight bounds for the mass flow $m$ and $\gamma$ variables, substantially reducing the feasible space of the OPGF problem. In addition,  Section~\ref{sec:relax_gap_quantification} introduces metrics to quantify the relaxation gap and discusses the impact of a nonzero gap on the solution.

\subsection{Deriving tight bounds on the mass flow}
\label{sec:bounds_mass_flow}
The relaxation-based models presented in Section~\ref{section:04_NGsol} require bounds $\underline{M}$ and $\overline{M}$ on the mass flow $m$ and bounds $\underline{\Gamma}$ and $\overline{\Gamma}$ on the auxiliary variable $\gamma$. Typically, the maximum pipeline flow is not part of the network data. Existing studies, e.g., \cite{June2022} and \cite{Schwele2019}, apply a large constant (Big-M) to the mass flow bounds without any physical relation. The mass flow in~\eqref{eq:momentum} is highest when the pressure gradient reaches its maximum, and the inertia term becomes zero (steady-state). In that case, tight mass flow bounds can be derived by applying the direction-dependent maximum feasible pressure drop along a pipeline, $\widehat{\Pi}^{+/-}$, to the Weymouth equation~\eqref{eq:weymouth}. Since the feasible pressure ranges might differ between adjacent nodes, the absolute value of $\underline{M}$ and $\overline{M}$ is generally not equivalent. Based on those values, tight bounds on the auxiliary variable $\gamma$ can be derived using its definition in~\eqref{eq:gamma_def}. 

\subsection{Quantifying the relaxation gap}
\label{sec:relax_gap_quantification}
The relaxations of~\eqref{eq:gamma_def} presented in Section~\ref{section:04_NGsol} will generally not be binding at the optimal solution, leading to physically infeasible operating points in practice.
This can lead to an overestimation of the linepack flexibility. 
Suppose there is a solution $(m_{ij,t}^*, \pi_{ij,t}^*, \gamma_{ij,t}^*)$, which satisfies~\eqref{eq:gamma_def}. Then solution $(m_{ij,t}^*, \pi_{ij,t}^{**}, \gamma_{ij,t}^*)$ with $\pi_{ij,t}^{**} > \pi_{ij,t}^{*}$ has a nonzero relaxation gap, i.e., the average pressure $\pi_{ij,t}^{**}$ is too high. This implies that the linepack, due to its proportional relationship with the average pressure by \eqref{eq:lp}, is also higher than it would be according to gas flow physics. Taking a look into the discretized PDEs~\eqref{eq:mass_d}--\eqref{eq:momentum_d}, there are two ways a nonzero relaxation gap reduces the system cost depending on whether the linepack is charged or discharged. When the pipeline is charged at time step $t$, a nonzero relaxation gap, i.e., $\pi_{ij,t}^{**} > \pi_{ij,t}^{*}$, implies that the net linepack charge is higher than according to the physics. On the other hand, when discharging a pipeline in time step $t$, choosing a nonzero relaxation gap in $t\!-\!1$, allows for a higher net discharging rate. 
Based on an illustrative case study, we demonstrate this in Section~\ref{subsection:results_choices}.
To assess and compare the quality of the solutions obtained with the relaxation-based models, we define the relative relaxation gap $\Phi_{ij,t}$ for each pipeline $(i,j)$ and time step $t$ as
\begin{equation}\label{eq:def_relaxation_gap}
    \Phi_{ij,t} = \frac{\gamma_{ij,t} - \frac{m_{ij,t}|m_{ij,t}|}{\pi_{ij,t}}}{\Gamma^{\mathrm{max}}_{ij}(z_{ij,t})},
\end{equation}
where $\Gamma^{\mathrm{max}}_{ij}(z_{ij})$ denotes the maximum or minimum value of $\gamma_{ij}$, respectively, depending on the flow direction $z_{ij,t}$. Based on the relative relaxation gap~\eqref{eq:def_relaxation_gap}, we define two metrics that aggregate the relative relaxation gap over the whole network by capturing the maximum absolute relaxation gap $\Phi^{\mathrm{\infty}}$ and the root mean squared (RMS) relaxation gap $\Phi^{\mathrm{RMS}}$, respectively:
\noindent\begin{minipage}{.4\linewidth}
\vspace{-16pt}
\begin{equation}\label{eq:gap_inf}
  \Phi^{\infty} = \lVert \mathbf{\Phi} \rVert_{\infty},
\end{equation}
\end{minipage}%
\begin{minipage}{.6\linewidth}
\begin{equation}\label{eq:gap_rms}
  \Phi^{\mathrm{RMS}} = \frac{1}{\sqrt{|\mathcal{T}||\mathcal{P}|}} \lVert \mathbf{\Phi} \rVert_{2},
\end{equation}
\vspace{3pt}
\end{minipage}
where $\mathbf{\Phi}$ is the vector that contains the relative relaxation gaps $\Phi_{ij,t}$ for all pipelines $(i,j)\in \mathcal{P}$ and time steps $t \in \mathcal{T}$. 
\section{Numerical Results} \label{section:results}
The optimization problems are solved on an Intel Xeon Processor 2650v4 with 256 GB RAM, 24 cores, and up to 2.20 GHz clock speed. The GitHub repository \cite{Online_Companion} contains all the input data and code implementation in JuMP v1.15.0 \cite{Lubin2023} for Julia v1.9.0 \cite{bezanson_julia_2017}. Except for $\mathbf{P}$-$\mathrm{NLP}$ solved by Ipopt v3.14.13 \cite{Waechter2005} with linear solver MA86 \cite{hogg2010indefinite}, all other models are solved with Gurobi v10.0.0 \cite{gurobi}.
Based on a stylized case study, we analyze the impact of modeling and solution choices on the linepack flexibility in Section~\ref{subsection:results_choices} and take a closer look at the flexibility provision in relaxation-based models in Section~\ref{sec:results_relaxations}. Section~\ref{large} compares solution choices' computational performance and accuracy based on a realistic large-scale case study.

\subsection{Linepack flexibility: Impact of modeling and solution choices}
\label{subsection:results_choices}
\begin{figure}[t]
    \centering
   \ctikzset{bipoles/length=0.8cm}

\resizebox{8.9cm}{!}{%
 \begin{circuitikz}[>=triangle 45]

        \begin{scope}[shift={(-1.5,-8)}, font=\large]
            \draw[thick] (-0.5,-0.5) rectangle (22.5,+0.5);

            \draw (0,0) node[draw, regular polygon, regular polygon sides=3, inner sep=3pt] {};
            \draw (0.3,0) node[anchor=west] {Gas source};
            \draw (3,0) node[circle, inner sep=3pt, draw] {};
            \draw (3.3,0) node[anchor=west] {Node};
            \draw (5,0) node[draw, rectangle, inner sep=5pt] {};
            \draw (5.3,0) node[anchor=west] {Gas load};

            \draw[thick] (7.5,-0.5) -- (7.5,+0.5);
            \draw[->] (8,-0.25) -- +(0,0.5); 
            \draw (8.25,0) node[anchor=west] {Electrical load};
            
            \draw[color = black] (11.75,-0.25) to [sV] (11.75,0.25) node[]{};
            \draw (12.25,0) node[anchor=west] {Non-GFPP};
            \draw[color = color_WF] (15.25,-0.25) to [sV] (15.25,0.25) node[]{};
            \draw (15.75,0) node[anchor=west] { Wind farm};
            \draw[color = color_GFPP] (18.75,-0.25) to [sV] (18.75,0.25) node[]{};
            \draw (19.25,0) node[anchor=west] {GFPP};
            \draw [ultra thick] (21,-0.25) -- (21,0.25);
            \draw (21.25,0) node[anchor=west] {Bus};
                    
        \end{scope}

        \scalebox{1.7}{  
        \begin{scope} [shift={(-1.5,-11.5)}]
    \draw [ultra thick] (2,9) node[anchor=south, rotate = 0]{\Large$n_{1}$} -- (2,8);
    \draw [ultra thick] (5,9) node[anchor=south, rotate = 0]{\Large$n_{2}$} -- (5,8);
    \draw [ultra thick] (3,10) node[anchor=east, rotate = 0]{\Large$n_{3}$} -- (4,10);
    \draw[->] (3.5,10) -- +(0,0.75); 
    \draw[->] (2,8.2) -- +(-0.75,0); 
    \draw (2,8.2) -- (5,8.2);  
    \draw (2,8.8) -- (2.25,8.8) -- (3.2,9.75) -- (3.2,10);  
    \draw (5,8.8) -- (4.75,8.8) -- (3.8,9.75) -- (3.8,10);  

    \draw (1.4,9.05)  to [sV] (1.4,8.55) node[anchor= north east, rotate=0, yshift = 15, xshift = -8]{\Large$g_1$}; 
    \draw (1.65,8.8)--(2,8.8);

    \draw[color = color_GFPP] (5.65,8.45)   to [sV] (5.65,7.95);
    \draw[color = color_GFPP] (5.025,8.2)--(5.4,8.2);
    \node[rotate=0] at (6.4,8.2) {\Large$\textcolor{color_GFPP}{g_2}$};

    \node[circle, fill=white, inner sep=0pt] (gfppP) at (5.65, 7.68) {};

    \draw [color = color_WF] (5.65,9.025)  to [sV] (5.65,8.575);
    \node[rotate=0] at (6.4,8.8) {\Large$\textcolor{color_WF}{w_1}$};
    \draw [color = color_WF] (5,8.8)--(5.4,8.8);

    \begin{scope}[shift={(-2,0.0)}]
        \pgfmathsetmacro{\s}{0.08}
        \node[draw, regular polygon, regular polygon sides=3, inner sep=3pt, label=above:\Large$i_{1}$] (i1) at (10,9.4) {};
        \node[draw, circle, inner sep=3pt, label=above:\Large$i_{2}$] (i2) at (13,9.4) {};
        \node[draw, regular polygon, regular polygon sides=3, inner sep=3pt, label=above:\Large$i_{3}$] (i3) at (15,9.4) {};
        \node[draw, rectangle, inner sep=5pt, label=right:\Large$i_{4}$] (i4) at (13,8.2) {};
    
        \draw (i1) -- (i2) node[midway,below] {\Large$p_{1}$};
        \draw (i2) -- (i3) node[midway,below] {\Large$p_{2}$};
        \draw (i2) -- (i4) node[midway,left] {\Large$p_{3}$};
            \node[circle, fill=white, inner sep=0pt] (gfppG) at (13, 7.68) {};
    \end{scope}

    \draw [dashed,color = color_GFPP] (i4) -- (gfppG) -- (gfppP) -- (5.65, 7.95);
    \end{scope}
    }
    \end{circuitikz}
}
   \vspace{-0.4cm}
\caption{Case Study A: Schematic diagram of the integrated power-gas network.}
\label{fig:small_case}
\figspace
\vspace{0.25cm}
\end{figure}
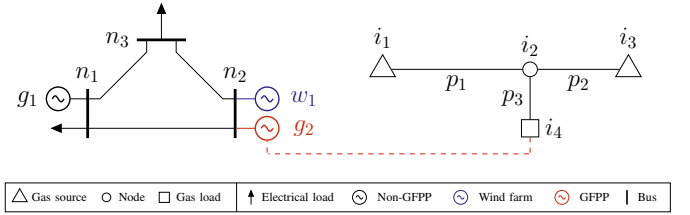

\begin{figure}[t]
    \centering
    \includegraphics[width=0.8\columnwidth]{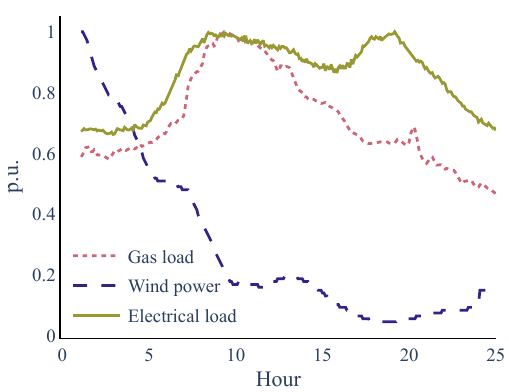}
    \caption{Gas load, wind power generation, and electricity load profiles in a \qty{5}{\minute} time resolution for an illustrative day \cite{energinet}.}
    \label{fig:profiles_day}
    \figspace
\end{figure}

We first consider Case Study A, a stylized $4$-node gas system connected to a $3$-bus power system, as depicted in Fig.~\ref{fig:small_case}. There is no compressor in this case study. The total capacity of two thermal power generators $g_1$ and $g_2$ equals the aggregated peak load of \qty{1500}{\mega\watt}. The expensive gas-fired generator $g_2$ is mainly used to balance fluctuations in the power supply of wind farm $w_1$, which has an installed capacity of \qty{750}{\mega\watt}. The normalized gas load, wind power generation, and electricity load profiles for an illustrative day are shown in Fig.~\ref{fig:profiles_day}, based on a \qty{5}{\minute} resolution \cite{energinet}. The peak gas load is \qty{80}{\kilo\gram\per\second}, excluding the gas demand of $g_2$. The gas and electricity loads peak at hours $9$ to $11$ when wind power generation drops significantly.
If the wind availability is low and not enough gas can be transported to $g_2$, then some of the electrical load must be shed.

We compare the flexibility provision obtained from the OPGF problem~\eqref{eq:problem_p_s} for the $\mathrm{DY}$, $\mathrm{QD}$, and $\mathrm{ST}$ models combined with a \qty{1}{\hour} and \qty{15}{\minute} time discretization, and solved with $\mathrm{NLP}$ and $\mathrm{MISOCP}$ methods. We use the original pipeline length to discretize in space dimension for illustration purposes. To ensure a transparent comparison, we enforce that the initial linepack for each pipeline must be restored at the end of the time horizon for the $\mathrm{DY}$ and $\mathrm{QD}$ models. We do not consider a restoration in the $\mathrm{ST}$ model as the intertemporal change in linepack is assumed to be zero (see Section~\ref{subsection:02_steadystate}).

\begin{figure*}
    \centering
    \includegraphics[width=1\textwidth]{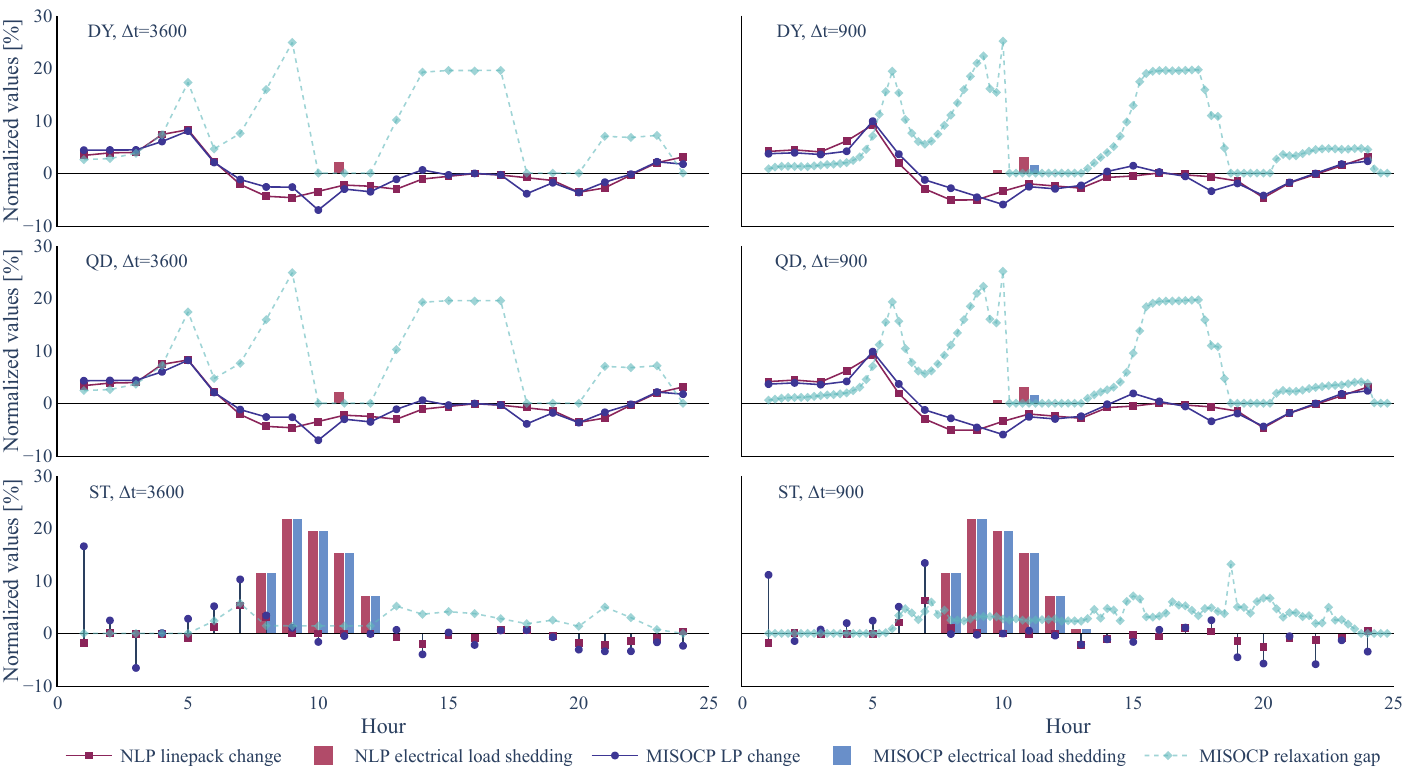}
    \caption{Case Study A: Change of linepack in pipeline $p_2$ and total electrical load shedding for the $\mathrm{DY}$ (first row), $\mathrm{QD}$ (second row), and $\mathrm{ST}$ (third row) models with a time discretization $\Delta t$ of \qty{1}{\hour} (first column) and  \qty{15}{\minute} (second column). Red and blue colors refer to the solution of the $\mathrm{NLP}$ and $\mathrm{MISOCP}$ models, respectively. For the latter, the relaxation gap~\eqref{eq:def_relaxation_gap} is shown in light blue. For the $\mathrm{ST}$ model, the linepack change is denoted differently from the $\mathrm{DY}$ and $\mathrm{QD}$ models (i.e., spike instead of continuous line) to indicate that the intertemporal linepack change cannot be exploited.
    The change of linepack is normalized by the maximum feasible linepack of pipeline $p_2$. The electrical load shedding is the share of the total hourly demand. The change in linepack and load shedding for the models with \qty{15}{\minute} resolution are aggregated over each hour.}
    \label{fig:coarse}
    \figspace
\end{figure*}

Fig.~\ref{fig:coarse} shows the optimal values for the change in linepack and relaxation gap for pipeline $p_2$ and the total electrical load shedding. We first focus on the $\mathrm{DY}$ model with a time discretization of \qty{1}{\hour} (top left). 
The pipeline $p_2$ is charged at the beginning and the end of the day and discharged in the middle to meet the high demand of the gas loads and GFPP $g_2$. For the $\mathrm{DY}$-$\mathrm{NLP}$ model, the amount of gas that can be transported to $g_2$ is insufficient to compensate for the reduced wind power generation, and therefore electrical load in hour $11$ is partially curtailed. In the $\mathrm{DY}$-$\mathrm{MISOCP}$ model, the significant relaxation gap in hour $9$ leads to flexibility overestimation, letting the linepack be discharged at a higher but physically infeasible rate in hour $10$. The discharged gas arrives at node $i_4$ in hour $11$. Hence, the electrical load curtailment is avoided. A similar observation can be made in hour $18$. We will further discuss the overestimation of flexibility in relaxation-based models in Section~\ref{sec:relax_gap_quantification}.

Looking at the first two rows, the $\mathrm{QD}$ and $\mathrm{DY}$ models achieve very similar linepack profiles, supporting the results found in \cite{Hennings2021} on the negligible impact of the inertia term. In the $\mathrm{ST}$ model, the linepack flexibility cannot be exploited. Therefore the electrical load shedding is inevitable in both $\mathrm{ST}$-$\mathrm{NLP}$ and $\mathrm{ST}$-$\mathrm{MISOCP}$ models. In the $\mathrm{ST}$-$\mathrm{NLP}$ model, \qty{1097}{\mega\watt\hour} of electricity are curtailed compared to only \qty{31}{\mega\watt\hour} in $\mathrm{DY}$-$\mathrm{NLP}$, demonstrating that neglecting the linepack change term in \eqref{eq:mass} does not allow to extract any flexibility from the gas network.
Both solution choices result in the same load-shedding decisions for the $\mathrm{ST}$ model, irrespective of the relaxation gap in the $\mathrm{ST}$-$\mathrm{MISOCP}$ model. Since the change in linepack is assumed to be zero in the $\mathrm{ST}$ model, the relaxation gap does not impact the results.\footnote{It is proven in \cite{Sing_Kekatos} that under certain assumptions, an optimal solution with zero relaxation gap exists for the $\mathrm{ST}$-$\mathrm{MISOCP}$ model.}

The right column in Fig.~\ref{fig:coarse} shows the modeling results for a time discretization of \qty{15}{\minute} instead of \qty{1}{\hour}, which increases the approximation accuracy of the PDEs.
This seems to reduce the utilization of linepack flexibility as the electrical load shedding increases. Even in the $\mathrm{DY}$-$\mathrm{MISOCP}$ model, the physically infeasible high discharge rate at hour $9$ is insufficient to avoid curtailment. Although the difference between $\mathrm{DY}$ and $\mathrm{QD}$ models is still negligible, it slightly increases with the finer time discretization from \qty{0.72}{\percent} to \qty{0.91}{\percent}. 
This is in line with the findings presented in \cite{Hennings2021}, which show that the effect of the inertia term increases with finer time discretization.

In contrast to the $\mathrm{QD}$ and $\mathrm{DY}$ models, time discretization does not influence the approximation accuracy of the PDEs for the $\mathrm{ST}$ model. A small impact still arises from considering more refined load profiles, resulting in a slight increase in load shedding. As we show in Section~$\mathrm{OC.2}$ of the online companion \cite{Online_Companion}, a coarser time discretization does not capture the extrema of time series happening on shorter time scales, e.g., \qty{5}{\min}. A more refined time discretization in component scheduling increases the decisions' accuracy and flexibility potential. Therefore, we will only use the $\mathrm{DY}$ model with \qty{15}{\min} resolution hereafter.

\begin{figure}
    \centering
    \includegraphics[width=0.8\columnwidth]{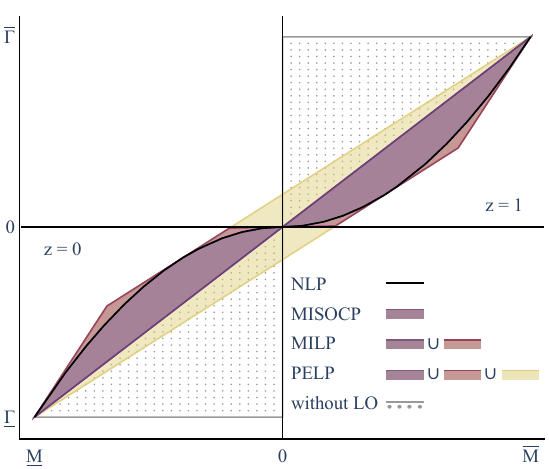}
    \caption{Comparison of the feasible regions defined by  different solution choices $\mathbf{S}$ for an illustrative pipeline and fixed pressure $\pi = \widehat{\Pi}$. The feasible regions of $\mathrm{MILP}$ and $\mathrm{MISOCP}$ are separated by the binary variable $z$ indicating the flow direction. The grey dotted area shows the extension of the feasible regions of $\mathrm{MILP}$ and $\mathrm{MISOCP}$ without the linear overestimator (LO)~\eqref{eq:LO_def}. For illustration purposes, a reduced number of halfspaces is shown for $\mathrm{PELP}$ and $\mathrm{MILP}$. All relaxations include the feasible region defined by the nonconvex constraint $\mathrm{NLP}$.}
    \label{fig:relaxation_comparison}
    \figspace
\end{figure}

\subsection{Linepack flexibility in relaxation-based models}
\vspace{2mm}
\label{sec:results_relaxations}
After observing a large influence of the relaxation gap in the $\mathrm{P}$-$\mathrm{MISOCP}$ model on the flexibility provision, this section further analyzes various relaxation-based solution choices $\mathbf{S}$. Let us take a closer look at their feasible regions $\mathcal{F}^{\mathbf{S}}$, depicted in Fig.~\ref{fig:relaxation_comparison}. The original nonconvex constraint~\eqref{eq:gamma_def} is shown in black. The feasible region of $\mathrm{MISOCP}$ is illustrated in purple and separated by the binary variable $z$, indicating the gas flow direction. Solution choice $\mathrm{MILP}$ constitutes an outer linear relaxation of $\mathrm{MISOCP}$, whose feasible region is the union of the red and purple areas. The feasible region of $\mathrm{PELP}$ additionally contains the yellow area. It is, therefore, a relaxation of $\mathrm{MILP}$. The dotted grey area shows the extension of the feasible region when the linear overestimator~\eqref{eq:LO_def} is excluded from $\mathrm{MISOCP}$ and $\mathrm{MILP}$. In this case, there exist points that are feasible for $\mathrm{PELP}$ but not for $\mathrm{MILP}$, and vice versa.

We now compare the flexibility results from their corresponding optimization problems. We again use the stylized Case Study A, except that we decrease the peak gas load to \qty{77.5}{\kilo\gram\per\second} to avoid the extreme case of load shedding. In the literature, e.g., \cite{Schwele2019}, flexibility provision from the gas to the power system is often quantified using the total absolute change in linepack $\Xi$ as
\begin{equation}
  \Xi = \sum_{(i,j) \in \mathcal{P}}\sum_{t \in \mathcal{T}} \lvert h_{ij,t} - h_{ij,t-1} \rvert,
\end{equation}
which shows how much the linepack is used as an intertemporal storage during a specific period $\mathcal{T}$. Table~\ref{tab:model_comparison} gives the results of the metrics to measure the relaxation gap $\Phi^{\infty} $, and $\Phi^{\mathrm{RMS}}$, the total absolute linepack change $\Xi$, as well as the total system operating cost, and computational time\footnote{We use the optimal solution of $\mathrm{MILP}$ as an initial guess for $\mathrm{MISOCP}$. Similarly, we use the optimal solution of $\mathrm{PELP}$ to warm start $\mathrm{SLP}$, as suggested in \cite{Mhanna2022}. We found both to reduce the computational time significantly and even improve the results in the case of $\mathrm{SLP}$. The aggregated solution times are given.} for the considered solution choices, including $\mathrm{MILP}$ and $\mathrm{MISOCP}$ without linear overestimator (w/o LO).

Models $\mathrm{DY}$-$\mathrm{NLP}$ to $\mathrm{DY}$-$\mathrm{PELP}$ are ordered with increasing feasible region $\mathcal{F}^{\mathbf{S}}$, as discussed before (see Fig.~\ref{fig:relaxation_comparison}). Model $\mathrm{DY}$-$\mathrm{NLP}$ is used as a reference for the total absolute linepack change and the total system operating cost.

Models $\mathrm{DY}$-$\mathrm{NLP}$ and $\mathrm{DY}$-$\mathrm{SLP}$ obtain the same solution (up to tolerances) with a zero relaxation gap, although $\mathrm{DY}$-$\mathrm{NLP}$ is solved faster. Model $\mathrm{DY}$-$\mathrm{MISOCP}$ finds a slightly better solution in terms of cost but frequently violates the gas flow physics, resulting in a root mean squared relaxation gap of \qty{6.12}{\percent}. Interestingly, the total absolute linepack change is \qty{0.09}{\percent} lower than for model $\mathrm{DY}$-$\mathrm{NLP}$, supporting the finding in Section~\ref{subsection:results_choices}, that the rate of linepack charge and discharge has a larger impact on flexibility than the total absolute change. In contrast to that, $\mathrm{DY}$-$\mathrm{MILP}$ finds a slightly better solution but utilizes the linepack nearly \qty{13}{\percent} more according to metric~$\Xi$. Both $\mathrm{DY}$-$\mathrm{MILP}$ and $\mathrm{DY}$-$\mathrm{MISOCP}$ exhibit a tremendous increase in computational time compared to $\mathrm{DY}$-$\mathrm{NLP}$ and $\mathrm{DY}$-$\mathrm{SLP}$. The model based on linear relaxation, i.e., $\mathrm{DY}$-$\mathrm{PELP}$, violates the gas flow physics at a $\Phi^{\mathrm{RMS}}$ of \qty{10.37}{\percent}, i.e., only slightly higher than $\mathrm{DY}$-$\mathrm{MISOCP}$. However, due to the absence of binary variables or nonconvex constraints, the computational time of $\mathrm{DY}$-$\mathrm{PELP}$ is much less than that of all other models.

Overall, the total absolute linepack change results show no causal relationship between a higher linepack utilization $\Xi$ and a reduction in total system operating cost, as indicated by \cite{Schwele2019}. In contrast to, e.g., battery storage, the linepack of the gas network features an additional geographical component, which restricts the degree of freedom from the intertemporal linepack storage.

The results in Table~\ref{tab:model_comparison} highlight that the relaxation gap, i.e., the violation of the gas flow physics, of solution choices $\mathrm{MILP}$ and $\mathrm{MISOCP}$ can be nearly halved by applying the linear overestimator~\eqref{eq:LO_def}. At the same time, it decreases the computational time. While this is not necessarily generalizable, including the linear overestimator seems highly favorable.

\begin{table}[]
    \centering
    \caption{Case Study A: Comparison of solution choices $\mathrm{S}$ for the $\mathrm{DY}$ model and $\Delta t = 900$. The total absolute linepack change and system operating cost are expressed relative to model~$\mathrm{NLP}$.}
    \begin{tabular}{l|rrrrr}
     \multicolumn{1}{c|}{\multirow{2}{*}{Model}} &  \multicolumn{1}{c}{$\Phi^{\mathrm{inf}}$} & \multicolumn{1}{c}{$\Phi^{\mathrm{RMS}}$} & \multicolumn{1}{c}{$\Xi$} & \multicolumn{1}{c}{Cost} & \multicolumn{1}{c}{Time} \\ 
     & \multicolumn{1}{c}{[\%]} & \multicolumn{1}{c}{[\%]} & \multicolumn{1}{c}{[\%]} & \multicolumn{1}{c}{[\%]} & \multicolumn{1}{c}{[s]} \\
    \hline
    NLP        & 0.00 & 0.00 & 0.00 & 0.00 & 0.84\Tstrut\\
    SLP          & 0.00 & 0.00 & 0.00 & 0.00 & 6.79\\
    MISOCP   & 24.40 & 6.12 & -0.09 & -0.73 & 167.81\\ 
    MILP   & 27.76 & 7.87 & +12.9 & -0.82 & 157.29\\
    PELP       & 32.09 & 10.37 & +8.28 & -0.94 & 0.06 \\
    \hline
    MISOCP w/o LO     & 43.43 & 12.12 & +16.22 & -0.85 & 415.59\Tstrut \\ 
    MILP w/o LO       & 65.15 & 15.62 & +30.44 & -0.93 & 283.10\\
    \hline
    \end{tabular}
    \label{tab:model_comparison}
    \figspace
    \vspace{2mm}
\end{table}

Comparing the models without linear overestimator to $\mathrm{DY}$-$\mathrm{PELP}$, it can be noted that they experience a much higher violation of the gas flow physics in terms of both relaxation gap metrics, but at the same time a lower cost reduction. This indicates that the feasible area of the solution choice $\mathrm{PELP}$ shown in yellow in the second and fourth quadrant, leads to more flexibility than the neglection of the linear overestimator. Operating points in that area include gas flows opposite to the pressure gradient.

\subsection{Solution choices for large systems} \label{large}
To analyze the impact of various solution choices on a realistic large-scale integrated power and gas system, we consider Case Study B, which is a modified version of the GasLib $40$-node gas network \cite{GasLib} connected to the modified $24$-bus IEEE RTS power system in \cite{ordoudis2016updated}. Section~$\mathrm{OC.5}$ of the online companion \cite{Online_Companion} describes the modifications to the gas network in detail. We use the same profiles shown in Fig.~\ref{fig:profiles_day} for electrical loads, gas loads, and wind farms, varying their peak load and installed capacity.
The data can be found in the GitHub repository \cite{Online_Companion}. A schematic diagram of the integrated system is shown in Fig.~\ref{fig:large_case}.

\definecolor{color_blue_1}{rgb}{0.267004,0.004874,0.329415}
\definecolor{color_blue_2}{rgb}{0.229739,0.322361,0.545706}
\definecolor{color_blue_3}{rgb}{0.128729,0.563265,0.551229}
\definecolor{color_blue_4}{rgb}{0.360741,0.785964,0.387814}
\definecolor{color_blue_5}{rgb}{0.993248,0.906157,0.143936}

\ctikzset{bipoles/length=0.85cm}

\def\legendx{0}
\def\legendy{0}

\def\boxheight{0.5cm}
\def\boxwidth{1cm}   

\newcommand{\mycolorbox}[1]{
    \fill[#1] ($(\legendx,\legendy)+(0,\theboxnum*\boxheight)$) rectangle ($(\legendx,\legendy)+(\boxwidth,\theboxnum*\boxheight+\boxheight))$;
    \addtocounter{boxnum}{1}
}
\def\ticklength{0.2cm}
\newcommand{\colortick}[2]{
    \draw ($(\legendx,\legendy)+(\boxwidth,#1*\boxheight)$) -- ($(\legendx,\legendy)+(\boxwidth+\ticklength,#1*\boxheight)$);
    \node[anchor=west] at ($(\legendx,\legendy)+(\boxwidth+\ticklength, #1*\boxheight)$) {#2};
}

\tikzset{
    compressor/.style={
        append after command={
            \pgfextra{
                \fill[black] ($(\tikzlastnode.north west)+(-0.1,0)$) -- ($(\tikzlastnode.east)+(0.1,0.3)$) -- ($(\tikzlastnode.east)+(0.1,-0.3)$) -- ($(\tikzlastnode.south west)+(-0.1,0)$) -- cycle;
}}}}

\tikzset{
    compressor90/.style={
        append after command={
            \pgfextra{
                \fill[black] ($(\tikzlastnode.north west)+(0,0.1)$) -- ($(\tikzlastnode.south)+(-0.3,-0.1)$) -- ($(\tikzlastnode.south)+(0.3,-0.1)$) -- ($(\tikzlastnode.north east)+(0,0.1)$) -- cycle;
}}}}

\tikzset{
    compressor180/.style={
        append after command={
            \pgfextra{
                \fill[black] ($(\tikzlastnode.north east)+(0.1,0)$) -- ($(\tikzlastnode.west)+(-0.1,0.3)$) -- ($(\tikzlastnode.west)+(-0.1,-0.3)$) -- ($(\tikzlastnode.south east)+(0.1,0)$) -- cycle;
}}}}

\begin{figure*}
    \centering

    \resizebox{18cm}{!}{%

     \begin{circuitikz}  

         \draw[thick] ($(\legendx,\legendy)+(-0.25,+0.25)$) rectangle ($(\legendx,\legendy)+(2.5,3.4)$);
        \draw ($(\legendx,\legendy)+(-0.125,3)$) node[anchor=west] {RMS [\%]};
        
        \fill[color_blue_1] ($(\legendx,\legendy)+(0,1*\boxheight)$) rectangle ($(\legendx,\legendy)+(\boxwidth,2*\boxheight)$);
        \fill[color_blue_2] ($(\legendx,\legendy)+(0,2*\boxheight)$) rectangle ($(\legendx,\legendy)+(\boxwidth,3*\boxheight)$);
        \fill[color_blue_3] ($(\legendx,\legendy)+(0,3*\boxheight)$) rectangle ($(\legendx,\legendy)+(\boxwidth,4*\boxheight)$);
        \fill[color_blue_4] ($(\legendx,\legendy)+(0,4*\boxheight)$) rectangle ($(\legendx,\legendy)+(\boxwidth,5*\boxheight)$);

        \draw ($(\legendx,\legendy)+(\boxwidth,2*\boxheight)$) -- ($(\legendx,\legendy)+(\boxwidth+\ticklength,2*\boxheight)$);
        \node[anchor=west] at ($(\legendx,\legendy)+(\boxwidth+\ticklength, 2*\boxheight)$) {2.5};
        \draw ($(\legendx,\legendy)+(\boxwidth,3*\boxheight)$) -- ($(\legendx,\legendy)+(\boxwidth+\ticklength,3*\boxheight)$);
        \node[anchor=west] at ($(\legendx,\legendy)+(\boxwidth+\ticklength, 3*\boxheight)$) {5};
        \draw ($(\legendx,\legendy)+(\boxwidth,4*\boxheight)$) -- ($(\legendx,\legendy)+(\boxwidth+\ticklength,4*\boxheight)$);
        \node[anchor=west] at ($(\legendx,\legendy)+(\boxwidth+\ticklength, 4*\boxheight)$) {7.5};

        \begin{scope}[shift={(2.5,-1.25)}]
            \draw[thick] (-0.5,-0.5) rectangle (28,+0.5);
        
            \draw (0,0) node[draw, regular polygon, regular polygon sides=3, inner sep=3pt] {};
            \draw (0.3,0) node[anchor=west] {Gas source};
            \draw (2.8,0) node[circle, inner sep=3pt, draw] {};
            \draw (3.1,0) node[anchor=west] {Node};
            \draw (5,0) node[draw, rectangle, inner sep=5pt] {};
            \draw (5.3,0) node[anchor=west] {Gas load};
            \draw (8,0) node[compressor180] {};
            \draw (8.3,0) node[anchor=west] {Compressor};

            \draw[thick] (11,-0.5) -- (11,+0.5);
            \draw[->] (12,-0.25) -- +(0,0.5); 
            \draw (12.3,0) node[anchor=west] {Electrical load};

            \draw[color = black] (16,-0.25) to [sV] (16,0.25) node[]{};
            \draw (16.5,0) node[anchor=west] {Non-GFPP};
            \draw[color = color_WF] (20,-0.25) to [sV] (20,0.25) node[]{};
            \draw (20.5,0) node[anchor=west] {Windgenerator};
            \draw[color = color_GFPP] (24,-0.25) to [sV] (24,0.25) node[]{};
            \draw (24.5,0) node[anchor=west] {GFPP};
            \draw [ultra thick] (26.3,-0.25) -- (26.3,0.25);
            \draw (26.6,0) node[anchor=west] {Bus};
            
        \end{scope}

        \foreach \type/\id/\no/\labelpos/\x/\y in {
            source/source_3/1/above/6/1.25,
            innode/innode_4/2/below/7.5/1.25,
            innode/innode_5/3/above/7.5/2.5,
            sink/sink_27/4/above/9/2.5,
            sink/sink_19/5/below/9/1.25,
            innode/innode_2/6/above/10.5/1.25,
            sink/sink_10/7/below/12/1.25,
            sink/sink_11/8/below/13.5/1.25,
            innode/innode_1/9/below/15/1.25,
            sink/sink_16/10/left/15/2.5,
            sink/sink_14/11/right/12/2.5,
            sink/sink_13/12/right/12/3.75,
            innode/innode_6/13/right/12/5,
            sink/sink_25/14/above right/12/6.25,
            source/source_1/15/right/15/7.5,
            sink/sink_3/16/right/15/6.25,
            sink/sink_23/17/left/15/5,
            innode/innode_8/18/above/13.5/6.25,
            source/source_2/19/above/9/10,
            innode/innode_7/20/above/10.5/10,
            sink/sink_29/21/right/12/10,
            sink/sink_28/22/above/12/11.25,
            sink/sink_2/23/above/13.5/8.75,
            sink/sink_15/24/left/12/7.5,
            sink/sink_20/25/below/10.5/5,
            sink/sink_4/26/below/9/6.25,
            sink/sink_8/27/above/9/3.75,
            sink/sink_17/28/below/7.5/5,
            sink/sink_5/29/below/6/5,
            sink/sink_7/30/below/4.5/5,
            sink/sink_24/31/below/3/5,
            sink/sink_21/32/below/1.5/5,
            sink/sink_12/33/below/0/5,
            sink/sink_26/34/above/9/7.5,
            sink/sink_9/35/above/7.5/7.5,
            sink/sink_18/36/above/6/7.5,
            sink/sink_6/37/above/4.5/7.5,
            sink/sink_22/38/above/2.25/7.5,
            sink/sink_1/39/above/0/7.5,
            subnode/subnode_1/40/above/10.8/1.25,
            subnode/subnode_2/41/above/11.1/1.25,
            subnode/subnode_3/42/above/11.4/1.25,
            subnode/subnode_4/43/above/11.7/1.25,
            subnode/subnode_5/44/above/12.5/9.58,
            subnode/subnode_6/45/above/13.0/9.17,
            subnode/subnode_7/46/above/11.7/6.0,
            subnode/subnode_8/47/above/11.4/5.75,
            subnode/subnode_9/48/above/11.1/5.5,
            subnode/subnode_10/49/above/10.8/5.25,
            subnode/subnode_11/50/above/5.5/7.5,
            subnode/subnode_12/51/above/5.0/7.5,
            subnode/subnode_13/52/above/3.5/5.0,
            subnode/subnode_14/53/above/4.0/5.0,
            subnode/subnode_15/54/above/12.0/1.56,
            subnode/subnode_16/55/above/12.0/1.88,
            subnode/subnode_17/56/above/12.0/2.19,
            subnode/subnode_18/57/above/12.0/6.88,
            subnode/subnode_19/58/above/9.75/4.38,
            subnode/subnode_20/59/above/12.75/1.25,
            subnode/subnode_21/60/above/9.0/7.08,
            subnode/subnode_22/61/above/9.0/6.67,
            subnode/subnode_23/62/above/3.75/7.5,
            subnode/subnode_24/63/above/3.0/7.5,
            subnode/subnode_25/64/above/1.12/7.5,
            subnode/subnode_26/65/above/11.5/6.46,
            subnode/subnode_27/66/above/11.0/6.67,
            subnode/subnode_28/67/above/10.5/6.88,
            subnode/subnode_29/68/above/10.0/7.08,
            subnode/subnode_30/69/above/9.5/7.29,
            subnode/subnode_31/70/above/11.0/10.0,
            subnode/subnode_32/71/above/11.5/10.0,
            subnode/subnode_33/72/above/9.0/1.88,
            subnode/subnode_34/73/above/15.0/1.46,
            subnode/subnode_35/74/above/15.0/1.67,
            subnode/subnode_36/75/above/15.0/1.88,
            subnode/subnode_37/76/above/15.0/2.08,
            subnode/subnode_38/77/above/15.0/2.29,
            subnode/subnode_39/78/above/6.75/5.0,
            subnode/subnode_40/79/above/8.25/7.5,
            subnode/subnode_41/80/above/7.88/1.25,
            subnode/subnode_42/81/above/8.25/1.25,
            subnode/subnode_43/82/above/8.62/1.25,
            subnode/subnode_44/83/above/13.12/6.25,
            subnode/subnode_45/84/above/12.75/6.25,
            subnode/subnode_46/85/above/12.38/6.25,
            subnode/subnode_47/86/above/12.0/10.62,
            subnode/subnode_48/87/above/8.0/2.5,
            subnode/subnode_49/88/above/8.5/2.5,
            subnode/subnode_50/89/above/12.0/4.38,
            subnode/subnode_51/90/above/12.0/9.17,
            subnode/subnode_52/91/above/12.0/8.33,
            subnode/subnode_53/92/above/9.75/5.62}
        {
            \ifthenelse{\equal{\type}{source}}{
                \node[draw, regular polygon, regular polygon sides=3, inner sep=3pt, label=\labelpos:\Large$i_{\no}$] (\id) at (\x,\y) {};
            }{
                \ifthenelse{\equal{\type}{innode}}{
                    \node[draw, circle, inner sep=3pt, label=\labelpos:\Large$i_{\no}$] (\id) at (\x,\y) {};
                }{
                    \ifthenelse{\equal{\type}{subnode}}{
                        \node[inner sep=0pt] (\id) at (\x,\y) {};
                    }{
                        \node[draw, rectangle, inner sep=5pt, label=\labelpos:\Large$i_{\no}$] (\id) at (\x,\y) {};
                    }
                }
            }
        }

        \draw[->, color=color_GFPP] (sink_16) -- +(0.75,0) node[below right] {\Large$g_1$}; 
        \draw[->, color=color_GFPP] (sink_10) -- +(0.75,-0.75) node[below right] {\Large$g_2$}; 
        \draw[->, color=color_GFPP] (sink_27) -- +(0.75,0) node[above right] {\Large$g_3$};
        \draw[->, color=color_GFPP] (sink_18) -- +(0,-0.75) node[below right] {\Large$g_{5/6}$}; 
        \draw[->, color=color_GFPP] (sink_26) -- +(0.75,0) node[above right] {\Large$g_7$}; 
        \draw[->, color=color_GFPP] (sink_2) -- +(0,-0.75) node[below right] {\Large$g_{10}$};
        \draw[->, color=color_GFPP] (sink_3) -- +(0.75,-0.75) node[below right] {\Large$g_{11/12}$};

        \foreach \start/\stop/\pipeNo/\color in {
            innode_4/innode_5/1/color_blue_1,
            sink_27/innode_5/2/color_blue_1,
            sink_27/sink_19/3/color_blue_2,
            innode_4/sink_19/4/color_blue_2,
            sink_10/innode_2/5/color_blue_1,
            sink_10/sink_11/6/color_blue_2,
            innode_1/sink_16/7/color_blue_1,
            sink_14/sink_10/8/color_blue_1,
            sink_13/sink_14/9/color_blue_2,
            innode_6/sink_13/10/color_blue_1,
            source_1/sink_3/11/color_blue_1,
            sink_3/sink_23/12/color_blue_2,
            sink_25/innode_8/13/color_blue_2,
            sink_29/innode_7/14/color_blue_1,
            sink_29/sink_28/15/color_blue_2,
            sink_29/sink_2/16/color_blue_1,
            sink_2/sink_15/17/color_blue_1,
            sink_15/sink_29/18/color_blue_1,
            sink_25/sink_15/19/color_blue_4,
            sink_25/sink_20/20/color_blue_2,
            sink_4/sink_20/21/color_blue_1,
            sink_17/sink_4/22/color_blue_1,
            sink_8/sink_20/23/color_blue_1,
            sink_17/sink_8/24/color_blue_1,
            sink_5/sink_17/25/color_blue_1,
            sink_7/sink_5/26/color_blue_1,
            sink_6/sink_7/27/color_blue_2,
            sink_7/sink_24/28/color_blue_3,
            sink_24/sink_21/29/color_blue_3,
            sink_21/sink_12/30/color_blue_1,
            sink_25/sink_26/31/color_blue_1,
            sink_26/sink_4/32/color_blue_1,
            sink_26/sink_9/33/color_blue_1,
            sink_9/sink_18/34/color_blue_1,
            sink_18/sink_6/35/color_blue_2,
            sink_6/sink_22/36/color_blue_1,
            sink_22/sink_1/37/color_blue_1,
            subnode_1/subnode_2/38/color_blue_1,
            subnode_2/subnode_3/39/color_blue_1,
            subnode_3/subnode_4/40/color_blue_1,
            subnode_4/sink_10/41/color_blue_1,
            subnode_5/subnode_6/42/color_blue_1,
            subnode_6/sink_2/43/color_blue_1,
            subnode_7/subnode_8/44/color_blue_1,
            subnode_8/subnode_9/45/color_blue_1,
            subnode_9/subnode_10/46/color_blue_1,
            subnode_10/sink_20/47/color_blue_1,
            subnode_11/subnode_12/48/color_blue_1,
            subnode_12/sink_6/49/color_blue_2,
            subnode_13/subnode_14/50/color_blue_1,
            subnode_14/sink_7/51/color_blue_1,
            subnode_15/subnode_16/52/color_blue_1,
            subnode_16/subnode_17/53/color_blue_1,
            subnode_17/sink_14/54/color_blue_2,
            subnode_18/sink_15/55/color_blue_2,
            subnode_19/sink_8/56/color_blue_1,
            subnode_20/sink_11/57/color_blue_3,
            subnode_21/subnode_22/58/color_blue_1,
            subnode_22/sink_4/59/color_blue_1,
            subnode_23/subnode_24/60/color_blue_1,
            subnode_24/sink_22/61/color_blue_1,
            subnode_25/sink_1/62/color_blue_1,
            subnode_26/subnode_27/63/color_blue_1,
            subnode_27/subnode_28/64/color_blue_1,
            subnode_28/subnode_29/65/color_blue_1,
            subnode_29/subnode_30/66/color_blue_1,
            subnode_30/sink_26/67/color_blue_1,
            subnode_31/subnode_32/68/color_blue_1,
            subnode_32/sink_29/69/color_blue_1,
            subnode_33/sink_19/70/color_blue_4,
            subnode_34/subnode_35/71/color_blue_1,
            subnode_35/subnode_36/72/color_blue_1,
            subnode_36/subnode_37/73/color_blue_1,
            subnode_37/subnode_38/74/color_blue_1,
            subnode_38/sink_16/75/color_blue_1,
            subnode_39/sink_5/76/color_blue_1,
            subnode_40/sink_9/77/color_blue_1,
            subnode_41/subnode_42/78/color_blue_1,
            subnode_42/subnode_43/79/color_blue_2,
            subnode_43/sink_19/80/color_blue_3,
            subnode_44/subnode_45/81/color_blue_1,
            subnode_45/subnode_46/82/color_blue_1,
            subnode_46/sink_25/83/color_blue_4,
            subnode_47/sink_28/84/color_blue_1,
            subnode_48/subnode_49/85/color_blue_1,
            subnode_49/sink_27/86/color_blue_1,
            subnode_50/innode_6/87/color_blue_3,
            subnode_51/subnode_52/88/color_blue_1,
            subnode_52/sink_15/89/color_blue_1,
            subnode_53/sink_4/90/color_blue_1}
        {
            \draw[color=\color, line width = 2.5, shorten >= -0.021cm, shorten <= -0.021cm] 
            (\start) -- (\stop); 
        }

        
        \foreach \id/\x/\y in {
            compressorStation_6/14.25/6.25
        }{
            \node[compressor] (\id) at (\x,\y) {};
        }
        
        \foreach \id/\x/\y in {
            compressorStation_2/14.25/1.25,
            compressorStation_3/9.75/1.25, 
            compressorStation_4/6.75/1.25, 
            compressorStation_5/9.75/10
        }{
            \node[compressor180] (\id) at (\x,\y) {};
        }
        
        \foreach \id/\label/\x/\y in {
            compressorStation_1/1/12/5.6
        }{
            \node[compressor90] (\id) at (\x,\y) {};
        }
        
        \foreach \from/\to in {
            compressorStation_1/sink_25,
            compressorStation_1/innode_6,
            compressorStation_2/innode_1,
            compressorStation_2/sink_11,
            compressorStation_3/innode_2,
            compressorStation_3/sink_19,
            compressorStation_4/innode_4,
            compressorStation_4/source_3,
            compressorStation_5/innode_7,
            compressorStation_5/source_2,
            compressorStation_6/innode_8,
            compressorStation_6/sink_3}
        {
            \draw[line width = 2] (\from) -- (\to);
        }

        \begin{scope}[shift={(17,0.65)}]

            \foreach \start/\name/\direction/\rot/\shift  in {
                {2,1}/n_{1}/north west/0/{-1,0},
                {10,1}/n_{2}/north west/0/{-1,0},
                {1,4.5}/n_{3}/east/0/{1,0},       
                {3.5,3.25}/n_{4}/east/0/{0,-1},      
                {7.5,3.25}/n_{5}/east/0/{0,-1},       
                {16,3}/n_{6}/north/0/{0,1},  
                {14,1}/n_{7}/north west/0/{-1,0},
                {16,1.5}/n_{8}/north/0/{0,1},    
                {9,4.5}/n_{9}/east/0/{1,0},        
                {13,4.5}/n_{10}/east/0/{1,0},     
                {9,6.25}/n_{11}/east/0/{1,0},
                {13,6.25}/n_{12}/east/0/{1,0},
                {16,6.75}/n_{13}/north/0/{0,1},
                {5,6.25}/n_{14}/east/0/{1,0},
                {1,8}/n_{15}/south east/0/{1,0},
                {6,8}/n_{16}/west/0/{-1,0}, 
                {1,9.75}/n_{17}/east/0/{1,0},
                {5,9.75}/n_{18}/east/0/{1,0},
                {9,8}/n_{19}/east/0/{1,0}, 
                {13,8}/n_{20}/east/0/{1,0}, 
                {9,9.75}/n_{21}/east/0/{1,0},
                {13,9.75}/n_{22}/east/0/{1,0},
                {16,9.25}/n_{23}/south/0/{0,-1},
                {1,6.25}/n_{24}/north east/0/{1,0}}
            {
                \node[anchor=\direction] (\name) at (\start) {\Large$\name$};
                \draw [ultra thick] (\start) -- ($(\start)+(\shift)$);
            }
    
            \draw[->] (1+0.75,1) -- +(0,-0.75); 
            \draw[->] (9+0.75,1) -- +(0,-0.75); 
            \draw[->] (1+0.5,4.5) -- +(0,-0.75); 
            \draw[->] (3.5,2.25+0.5) -- +(0.75,0); 
            \draw[->] (7.5,2.25+0.5) -- +(-0.75,0); 
            \draw[->] (16,3+0.5) -- +(0.75,0); 
            \draw[->] (13+0.75,1) -- +(0,-0.75); 
            \draw[->] (16,1.5+0.5) -- +(0.75,0); 
            \draw[->] (9+0.5,4.5) -- +(0,0.75); 
            \draw[->] (13+0.5,4.5) -- +(0,0.75); 
            \draw[->] (16,6.25+0.75) -- +(0.75,0); 
            \draw[->] (5+0.5,6.25) -- +(0,-0.75); 
            \draw[->] (1+0.25,8) -- +(0,0.75); 
            \draw[->] (5+0.75,8) -- +(0,-0.75); 
            \draw[->] (5+0.75,9.75) -- +(0,+0.75); 
            \draw[->] (9+0.5,8) -- +(0,-0.75); 
            \draw[->] (13+0.5,8) -- +(0,-0.75); 

            \draw (13.75,1) -- (13.75,1.25) -- (15.75,1.75) -- (16,1.75); 
            \draw (16,3.25) -- (15.75,3.25) -- (9.75,1.25) -- (9.75,1);   
            \draw (16,3.75) -- (15.75,3.75) -- (13.75,4.25) -- (13.75,4.5); 
            \draw (7.5,3) -- (7.75,3) -- (13.25,4.25) -- (13.25,4.5);    
            \draw (16,2.25) -- (15.75,2.25) -- (13.5,4.25) -- (13.5,4.5); 
            \draw (16,2) -- (15.75,2) -- (9.75,4.25) -- (9.75,4.5);   
            \draw (3.5,3) -- (3.75,3) -- (9.5,4) -- (9.5,4.5);       
            \draw (3.5,2.5) -- (3.75,2.5) -- (9.5,1.5) -- (9.5,1);     
            \draw (1.75,1) -- (1.75,1.25) -- (9.25,1.25) -- (9.25,1);    
            \draw (1.5,1) -- (1.5,1.5) -- (7.25,2.5) -- (7.5,2.5);      
            \draw (1.25,4.5) -- (1.25,4.25) -- (1.25,1.5) -- (1.25,1);     
            \draw (1.75,4.5) -- (1.75,4.25) -- (9.25,4.25) -- (9.25,4.5);     
            \draw (13.25,6.25) -- (13.25,6) -- (9.75,5) -- (9.75,4.5);  
            \draw (13.25,4.5) -- (13.25,4.75) -- (9.75,6) -- (9.75,6.25);  
            \draw (13.75,4.5) -- (13.75,6.25);  
            \draw (9.25,4.5) -- (9.25,6.25);  
            \draw (1.5,4.5) -- (1.5,6.25);  
            \draw (1.5,8) -- (1.5,6.25);  
            \draw (13.75,6.25) -- (13.75,6.5) -- (15.75,7) -- (16,7);  
            \draw (9.75,6.25) -- (9.75,6.5) -- (15.75,7.25) -- (16,7.25);  
            \draw (16,7.5) -- (15.75,7.5) -- (15.75,8.5) -- (16,8.5);  
            \draw (13.25,6.25) -- (13.25,6.5) -- (15.75,8.75) -- (16,8.75);  
            \draw (13.75,8) -- (13.75,8.25) -- (15.75,9) -- (16,9);  
            \draw (5.75,6.25) -- (5.75,6.5) -- (9.25,6.5) -- (9.25,6.25);  
            \draw (5.5,8) -- (5.5,6.25);  
            \draw (1.75,8) -- (1.75,8.25) -- (5.25,8.25) -- (5.25,8); 
            \draw (1.5,8) -- (1.5,8.5) -- (9.5,9) -- (9.5,9.75); 
            \draw (1.5,9.75) -- (1.5,9.25) -- (13.75,9.25) -- (13.75,9.75);  
            \draw (9.75,8) -- (9.75,8.25) -- (13.25,8.25) -- (13.25,8); 
            \draw (5.75,8) -- (5.75,8.25) -- (9.25,8.25) -- (9.25,8);  
            \draw (1.25,9.75) -- (1.25,9) -- (5.5,8.5) -- (5.5,8);  
            \draw (1.75,9.75) -- (1.75,9.5) -- (5.25,9.5) -- (5.25,9.75);  
            \draw (5.75,9.75) -- (5.75,9.5) -- (9.25,9.5) -- (9.25,9.75);  
            \draw (9.75,9.75) -- (9.75,9.5) -- (13.25,9.5) -- (13.25,9.75); 
    
            \draw[color = color_GFPP] (1.25,0.75) to [sV] (1.25,0.25) node[anchor=north east]{\Large$g_1$};
            \draw[color = color_GFPP] (1.25,1)--(1.25,0.75);
            \draw[color = color_GFPP] (9.25,0.75) to [sV] (9.25,0.25) node[anchor=north east]{\Large$g_2$};
            \draw[color = color_GFPP] (9.25,1)--(9.25,0.75);
            \draw[color = color_GFPP] (13.25,0.75) to [sV] (13.25,0.25) node[anchor=north east]{\Large$g_3$};
            \draw[color = color_GFPP] (13.25,1)--(13.25,0.75);
            \draw[color = black] (16.5,7.75) node[anchor = north west, rotate=0, xshift=10, yshift=3]{\Large $g_4$}  to [sV] (16.5,7.25) ;
            \draw[color = black] (16,7.5)--(16.25,7.5);
            \draw[color = color_GFPP] (0.75,7.75) to [sV] (0.75,7.25) node[anchor=north east, xshift=0, yshift=0]{\Large$g_{5/6}$};
            \draw[color = color_GFPP] (1.25,8)--(1.25,7.5)--(1,7.5);
            \draw[color = color_GFPP] (4.75,7.75) to [sV] (4.75,7.25) node[anchor=north east, xshift=0, yshift=0]{\Large$g_{7}$};
            \draw[color = color_GFPP] (5.25,8)--(5.25,7.5)--(5,7.5);
            \draw[color = color_GFPP] (16.5,8.75) node[anchor = north west, rotate=0, xshift=10, yshift=3]{\Large $g_{11/12}$}  to [sV] (16.5,8.25) ;
            \draw[color = color_GFPP] (16,8.5)--(16.25,8.5);
            \draw[color = black] (5.25,10.5) to [sV] (5.25,10) node[anchor=south east, yshift=12]{\Large$g_8$};
            \draw[color = black] (5.25,10)--(5.25,9.75);
            \draw[color = black] (9.25,10.5) to [sV] (9.25,10) node[anchor=south east, yshift=12]{\Large$g_9$};
            \draw[color = black] (9.25,10)--(9.25,9.75);
            \draw[color = color_GFPP] (13.25,10.5) to [sV] (13.25,10) node[anchor=south east, yshift=12]{\Large$g_{10}$};
            \draw[color = color_GFPP] (13.25,10)--(13.25,9.75);

            \draw[color = color_WF] (2.25,5.25) to [sV] (2.25,4.75) node[anchor=south west, xshift=0, yshift=12]{\Large$w_{1}$};
            \draw[color = color_WF] (1.75,4.5)--(1.75,5)--(2,5);
            \draw[color = color_WF] (8,2.75) to [sV] (8,2.25) node[anchor=north west, xshift=2, yshift=0]{\Large$w_{2}$};
            \draw[color = color_WF] (7.5,2.5)--(7.75,2.5);
            \draw[color = color_WF] (13.25,1.75) to [sV] (13.25,1.25) node[anchor=south west, xshift=5, yshift=10]{\Large$w_3$};
            \draw[color = color_WF] (13.25,1)--(13.25,1.25);
            \draw[color = color_WF] (2.25,7.75) to [sV] (2.25,7.25) node[anchor=north west, xshift=0, yshift=-2]{\Large$w_{4}$};
            \draw[color = color_WF] (1.75,8)--(1.75,7.5)--(2,7.5);
            \draw[color = color_WF] (10.5,10.5) to [sV] (10.5,10) node[anchor=south west, xshift=0, yshift=12]{\Large$w_{5}$};
            \draw[color = color_WF] (9.75,9.75)--(9.75,10.25)--(10.25,10.25);
        
        \end{scope}

            
    \end{circuitikz}
    }

\caption{Case study B: GasLib 40-node gas network connected to 24-bus IEEE RTS system. Pipeline colors indicate the value of the RMS relaxation gap of each discretized pipeline segment for the optimal solution of model $\mathrm{DY}$-$\mathrm{PELP}$.}
\label{fig:large_case}
\figspace
\end{figure*}
%

We consider the $\mathrm{DY}$ model only and apply a time and space discretization of \qty{15}{\minute} and \qty{15}{\kilo\meter}, respectively. Each model is run 10 times and the mean solution time is reported.
Except for solution choices $\mathrm{NLP}$, $\mathrm{SLP}$, and $\mathrm{PELP}$, no other method was able to solve the OPGF problem within $24$ hours.

The three models achieve the same solution regarding the total operating cost of the integrated system.
Model $\mathrm{DY}$-$\mathrm{NLP}$ is solved to local optimality in \qty{ 394.78}{\second}. Based on the stopping criteria in Appendix~\ref{sec:SLP_app}, Algorithm~$\mathrm{DY}$-$\mathrm{SLP}$ converges after 5 iterations in \qty{36.72}{\second}.
This confirms the finding in \cite{ConorThesis}, that $\mathrm{SLP}$ scales better than $\mathrm{NLP}$ in terms of solution time. The linear relaxation $\mathrm{DY}$-$\mathrm{PELP}$ is solved to global optimality in \qty{2.44}{\second}, but the solution violates the gas flow physics. The values for the relaxation gap metrics are \qty{8.38}{\percent} and \qty{2.17}{\percent} for $\Phi^{\infty}$ and $\Phi^{\mathrm{RMS}}$, respectively.

For Case Study B, the optimal solutions to $\mathrm{DY}$-$\mathrm{NLP}$ and $\mathrm{DY}$-$\mathrm{SLP}$ each include six changes of flow directions over all pipelines and time steps. On the contrary, the $\mathrm{DY}$-$\mathrm{PELP}$ model does not experience any change of flow direction owed to operating points that have opposite flow and pressure gradient directions, as explained in Section~\ref{sec:results_relaxations}. Those physically infeasible operating points significantly overestimate the flexibility provision from gas networks to power systems.

The different colors of the pipelines in Fig.~\ref{fig:large_case} illustrate each pipeline's RMS relaxation gap\footnote{This is done by calculating $\Phi^{\mathrm{RMS}}$ according to Equation~\eqref{eq:gap_rms} for each pipeline $(i,j)$ using the stacked relative relaxation gaps $\mathbf{\Phi}_{ij}$ and without normalizing by the square-root of the total number of pipelines.} for model $\mathrm{DY}$-$\mathrm{PELP}$.
Generally, it can be observed that the relaxation gap is, on average, higher around junctions. Suppose, for example, that gas flows from a junction into two pipelines with the same pressure gradient. Then, the relaxation gap can be utilized to have a proportionally lower flow on one of the connected branches. For solution choices $\mathrm{MILP}$ and $\mathrm{PELP}$, an over-proportionally high flow compared to the pressure gradient could also be feasible. However, since the pressure ranges are usually nonbinding, this rarely happens, as a larger pressure gradient could be chosen.

As the $\mathrm{DY}$-$\mathrm{PELP}$ model is a convex relaxation of $\mathrm{DY}$-$\mathrm{NLP}$, its solution constitutes a global lower bound to the OPGF problem, which is also pointed out in \cite{Mhanna2022}. Interestingly, even though the operational schedules obtained by the $\mathrm{DY}$-$\mathrm{PELP}$ model differ substantially from those by $\mathrm{DY}$-$\mathrm{NLP}$ and $\mathrm{DY}$-$\mathrm{SLP}$ models, the total operating cost of the integrated system is only slightly lower (less than 0.01\%). This indicates that both solution choices $\mathrm{NLP}$ and $\mathrm{SLP}$ find nearly globally optimal solutions, as pointed out also by other studies, such as \cite{Mhanna2022}. 
\section{Recommendations and future works}
\label{section:05_conclusion}
An accurate representation of the slow gas flow dynamics is crucial to capture the flexibility that gas networks can provide to power systems. A set of nonlinear and nonconvex PDEs governs these dynamics. This paper provided a unified PDE-based framework. In this framework, three independent decisions have to be taken, affecting the accuracy of the gas flow physics representation and the flexibility provision. We group these decisions into modeling and solution choices as illustrated in Fig.~\ref{fig:framework_choices}.
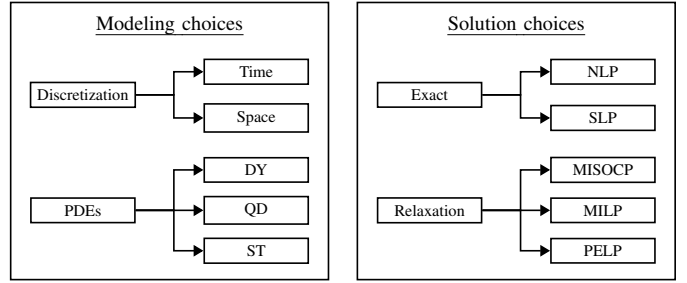
\begin{figure}
    \centering
    \resizebox{8.95cm}{!}{%
\begin{tikzpicture}[
    modelling/.style = {
        rectangle, draw = black, thick, anchor = center, minimum height = 0.3 cm, minimum width = 1.8 cm,
        font=\footnotesize,
    },
    solution/.style = {
        rectangle, draw = black, thick, anchor = center, minimum height = 0.3 cm, minimum width = 1.8 cm,
        font=\footnotesize,
    }
    ]

    
    \node [modelling] (m11) at (5,1) {Discretization};
    \node [modelling] (m12) at (5,-1) {PDEs};

    \node (mm11) at (6.5,1) {};
    \node (mm12) at (6.5,-1) {};
    
    \node [modelling] (m111) at (8,1.4) {Time};
    \node [modelling] (m112) at (8, 0.6) {Space};
    \node [modelling] (m121) at (8,-0.3) {DY};
    \node [modelling] (m122) at (8,-1) {QD};
    \node [modelling] (m123) at (8,-1.7) {ST};

    \draw[->,>=Triangle,thick] (m11.east) -- ([xshift=-0.05cm]mm11.east) |- (m111.west) {};
    \draw[->,>=Triangle,thick] (m11.east) -- ([xshift=-0.05cm]mm11.east) |- (m112.west) {};
    \draw[->,>=Triangle,thick] (m12.east) -- ([xshift=-0.05cm]mm12.east) |- (m121.west) {};
    \draw[->,>=Triangle,thick] (m12.east) -- ([xshift=-0.05cm]mm12.east) |- (m122.west) {};
    \draw[->,>=Triangle,thick] (m12.east) -- ([xshift=-0.05cm]mm12.east) |- (m123.west) {};

    \draw[thick] (3.75,-2.2) rectangle (9.25,2.6);
    \node at (6.5,2.2) {\uline{Modeling choices}};

    
    \node [solution] (m21) at ([xshift=6cm]m11) {Exact};
    \node [solution] (m22) at ([xshift=6cm]m12) {Relaxation};

    \node (mm21) at ([xshift=6cm]mm11) {};
    \node (mm22) at ([xshift=6cm]mm12) {};
    
    \node [solution] (m211) at ([xshift=6cm]m111) {NLP};
    \node [solution] (m212) at ([xshift=6cm]m112) {SLP};
    \node [solution] (m221) at ([xshift=6cm]m121) {MISOCP};
    \node [solution] (m222) at ([xshift=6cm]m122) {MILP};
    \node [solution] (m223) at ([xshift=6cm]m123) {PELP};

    \draw[thick] (9.75,-2.2) rectangle (15.25,2.6);
    \node at (12.5,2.2) {\uline{Solution choices}};

    \draw[->,>=Triangle,thick] (m21.east) -- ([xshift=-0.05cm]mm21.east) |- (m211.west) {};
    \draw[->,>=Triangle,thick] (m21.east) -- ([xshift=-0.05cm]mm21.east) |- (m212.west) {};
    \draw[->,>=Triangle,thick] (m22.east) -- ([xshift=-0.05cm]mm22.east) |- (m221.west) {};
    \draw[->,>=Triangle,thick] (m22.east) -- ([xshift=-0.05cm]mm22.east) |- (m222.west) {};
    \draw[->,>=Triangle,thick] (m22.east) -- ([xshift=-0.05cm]mm22.east) |- (m223.west) {};
\end{tikzpicture}
}
    \vspace{-0.8cm}
    \caption{Modeling and solution choices in the PDE-based gas flow framework.}
    \figspace
    \label{fig:framework_choices}
\end{figure}
The first decision is on the selection of a variant of the PDEs~\eqref{eq:mass}--\eqref{eq:momentum}, including the one that considers all terms ($\mathrm{DY}$ model), or the one that neglects the inertia term ($\mathrm{QD}$ model), or eventually the one that discards all temporal dynamics, including linepack flexibility ($\mathrm{ST}$ model). The second decision pertains to the time and space discretization that transforms the chosen PDE model into nonlinear and nonconvex algebraic equations. The finer the discretization is, the more accurate the approximation of the PDEs will be. Finally, the third decision corresponds to the solution choice to solve either the exact optimization problem ($\mathrm{NLP}$ or $\mathrm{SLP}$) or its relaxations ($\mathrm{MISOCP}$, $\mathrm{MILP}$, or $\mathrm{PELP}$). 

We showed how reformulations of the $\mathrm{QD}$ and $\mathrm{ST}$ models based on the Weymouth equation and different solution choices adopted in the literature fit in the proposed PDE-based framework. The framework is complemented by some general recommendations to tackle numerical challenges, tighten the feasible region of individual models, and quantify the quality of relaxation-based solution choices.

Based on this framework, we assessed the impact of various modeling and solution choices on the flexibility provision in the OPGF problem.
We found the $\mathrm{ST}$ model unsuitable for operational decisions, as it does not capture linepack flexibility. It may, however, be suitable for long-term planning decisions, which is outside of the scope of this analysis. The $\mathrm{QD}$ and $\mathrm{DY}$ models yield equivalent flexibility results as the influence of the inertia term is minimal. Refining the temporal discretization from \qty{1}{\hour} to \qty{15}{\minute} fosters an accurate estimation of the potential flexibility provision. Moving to markets, models, and data with finer temporal resolution improves the flexibility utilization. The proposed PDE-based framework allows for an application of relaxation-based solution choices to the $\mathrm{DY}$ model. Relaxation-based solution choices tend to overestimate the linepack flexibility. When the relaxation gap is nonzero, the linepack can be charged and discharged at a physically infeasible rate to avoid load shedding and the use of expensive gas sources. Our analysis showed that the relaxation-based solution choices $\mathrm{MISOCP}$ and $\mathrm{MILP}$ requiring binary variables for modeling the flow directions do not computationally scale well and deliver solutions significantly violating the gas flow physics. The violations and the computational complexity can be substantially reduced by including a linear overestimator. Solution choice $\mathrm{PELP}$ delivers slightly worse results regarding the relaxation gap but at a much lower computational cost, even for large case studies. We conclude that exact solution techniques, such as $\mathrm{NLP}$ and $\mathrm{SLP}$, should be preferred for operational decisions and assessment of linepack flexibility. Those can be complemented by $\mathrm{PELP}$, suitable for generating a global lower bound to the OPGF problem.
With increasing problem size, $\mathrm{SLP}$ seems to outperform $\mathrm{NLP}$ computationally while finding equally satisfactory solutions.
These findings confirm the results in \cite{Mhanna2022}.

Several studies have proposed detailed compressor models that capture their underlying nonlinearities and nonconvexities \cite{koch2015evaluating, rose2016computational}. Future research should focus on verifying the findings of this paper when including detailed compressor models in the proposed PDE-based framework. The nonlinear and nonconvex gas compression physics may improve the tightness of relaxation-based solution choices. Furthermore, future studies should investigate the impact of adopting a non-ideal equation of state \cite{9999472} on flexibility estimation.
The presented OPGF problem and associated solution choices, especially $\mathrm{NLP}$ and $\mathrm{SLP}$, should be extended to account for binary operating decisions in integrated power and gas systems, such as the unit commitment of conventional generators or gas valve states (open/closed). Finally, financial compensation for the contribution to linepack may incentivize the utilization of flexibility. This requires careful linepack pricing that considers the PDEs' spatial and temporal dynamics.

\appendices
\section{Optimal power and gas flow}
\label{section:appendix_OPT}
In the following, we introduce the full formulation of the optimal power-gas flow problem \eqref{eq:OPGF_opt}. Let $t \in \mathcal{T}$ denote the set of time steps. For the gas network, sets $\mathcal{S}$, $\mathcal{D}$, $\mathcal{P}$, $\mathcal{C}$ represent gas suppliers, gas loads, pipelines, and compressors, respectively. Gas nodes are indicated by $i,j \in \mathcal{I}$. For the power system, $\mathcal{K}$, $\mathcal{G}\subseteq \mathcal{K}$, $\mathcal{W}$, $\mathcal{R}$, $\mathcal{L}$, denote dispatchable generators, gas-fired power plants (GFPPs), wind farms, electrical loads, and lines. Power buses are represented by $n,m \in \mathcal{N}$. The sets $\Psi^{(.)}_i$ and $\Psi^{(.)}_n$ denote, respectively, the assets located at gas network $i$ and at power bus $n$. Gas nodes connected to node $i$ and power buses connected to bus $n$ are denoted by $\Omega_i$ and $\Omega_n$, respectively.

\subsection{Objective function}
The objective function \eqref{eq:ob_OPGF} minimizes the total system operating cost over the time steps $t$ of length $\Delta t$, as the sum of gas supplies $q^{\mathrm{S}}_{s,t}$ and power generation $p_{k,t}$ from non-GFPPs. Quadratic cost functions are assumed for gas suppliers $s$ and generators. 
For power, $A_k^{\mathrm{L}}$  and $A_k^{\mathrm{Q}}$ are the linear and quadratic cost coefficients, respectively. For gas, $A_s^{\mathrm{L}}$  and $A_s^{\mathrm{Q}}$ are the linear and quadratic cost coefficients, respectively. 
Gas and electrical load shedding, i.e., $q^{\mathrm{O}}_{d,t}$ and $p^{\mathrm{O}}_{r,t}$, are allowed at the prices $C^{\mathrm{sh, G}}$ and $C^{\mathrm{sh, P}}$, respectively.
\begin{align}
\begin{split}
 \underset{\mathbf{w}, \mathbf{v}, \mathbf{x}, \mathbf{y}} {\mathrm{min}}  &\quad \frac{\Delta t}{3600} \sum_{t \in \mathcal{T}} \Biggr[
 \sum_{s \in \mathcal{S}} (A_s^{\mathrm{Q}} (q^{\mathrm{S}}_{s,t})^2 + A_s^{\mathrm{L}} q^{\mathrm{S}}_{s,t}) +  \sum_{d \in \mathcal{D}} C^{\mathrm{sh, G}} q^{\mathrm{O}}_{d,t} \\
&+ \sum_{k \in (\mathcal{K} \setminus \mathcal{G})} (A_k^{\mathrm{Q}} p_{k,t}^2 + A_k^{\mathrm{L}} p_{k,t}) + \sum_{r \in \mathcal{R}} C^{\mathrm{sh, P}} p^{\mathrm{O}}_{r,t} \Biggl]. \label{eq:ob_OPGF}
\end{split}
\end{align}
%
\subsection{Power system constraints}
The power system constraints are enforced for each time step $t \in \mathcal{T}$ as follows:
\begin{subequations}
\begin{align}
& \underline{P}_k\leq p^{\mathrm{G}}_{k,t} \leq \overline{P}_k, & \forall\, & k \in \mathcal{K},\label{eq:el_p_lim_gen} \\
& 0 \leq p^{\mathrm{W}}_{w,t} \leq P^{\mathrm{W}}_{w,t}, & \forall\, & w \in \mathcal{W}, \label{eq:el_p_lim_wind}\\
& 0\leq p^{\mathrm{O}}_{r,t} \leq P^{\mathrm{D}}_{r,t}, & \forall\, & r \in \mathcal{R}, \label{eq:el_curtailment}\\
& f_{nm,t}  = B_{nm} (\theta_{n,t} -\theta_{m,t}) , & \forall\, & (n,m) \in \mathcal{L}, \label{eq:el_p_line}\\
& -F_{nm}  \leq f_{nm,t} \leq F_{nm} , & \forall\, & (n,m) \in \mathcal{L}, \label{eq:el_p_lim_line}\\
&  \theta_{n,t} = 0 , & n & =  n^{\mathrm{ref}}, \label{eq:el_theta_ref} 
\end{align}
\end{subequations}
\vspace{-0.5cm}
\begin{align}
\begin{split}\label{eq:el_balance}
\hspace{0.5em} \sum_{k \in \Psi^{\mathcal{K}}_n} & p^{\mathrm{G}}_{k,t} + \sum_{w \in \Psi^{\mathcal{W}}_n} p^{\mathrm{W}}_{w,t} -  \sum_{m \in \Omega_n} f_{nm,t} \\
&- \sum_{r \in \Psi^{\mathcal{R}}_n} (P^{D}_{r,t} - p^{\mathrm{O}}_{r,t})  = 0, \qquad \quad \forall\, n \in \mathcal{N}. 
\end{split}
\end{align}
Constraint \eqref{eq:el_p_lim_gen} ensures that dispatchable generators operate between their minimum and maximum limits, $\underline{P}_k$ and $\overline{P}_k$. In~\eqref{eq:el_p_lim_wind}, the wind dispatch $p^{\mathrm{W}}_{w,t}$ is upper bounded by the forecasted wind production $P^{\mathrm{W}}_{w,t}$. Constraint~\eqref{eq:el_curtailment} states that electrical load shedding should not exceed the electrical load $P^{\mathrm{D}}_{r,t}$. Constraint \eqref{eq:el_p_line} is the DC power flow approximation, where $f_{nm,t}$ is the line power flow, $\theta_{n,t}$ and $\theta_{m,t}$ the nodal voltage angles, and $B_{nm}$ is the line susceptance. Constraint \eqref{eq:el_theta_ref} enforces the power flow to be within the line capacity limit $F_{nm}$. Constraint \eqref{eq:el_theta_ref} defines the reference voltage bus. Finally, \eqref{eq:el_balance} ensures the nodal power balance between generation and demand.

\subsection{Gas system constraints}
The gas system constraints are enforced for each time step $t \in \mathcal{T}$ as follows:
\begin{subequations}
\begin{align}
& \underline{Q}_s\leq q^{\mathrm{S}}_{s,t} \leq \overline{Q}_s, & \forall\, & s \in \mathcal{S}, \label{eq:gas_q_lim_source} \\
& \underline{\Pi}_i\leq \pi_{i,t} \leq \overline{\Pi}_i, & \forall\, & i \in \mathcal{I}, \label{eq:gas_pres_lim} \\
& \pi_{i,t} = \Pi_i^{\mathrm{S}}, & \forall\, & i \in \mathcal{I}^{\mathrm{S}}, \label{eq:gas_pres_lim_s} \\
& 0\leq q^{\mathrm{O}}_{d,t} \leq Q^{\mathrm{D}}_{d,t}, & \forall\, & d \in \mathcal{D}, \label{eq:gas_curtailment}\\
& q^{\mathrm{C}}_{ij,t} \geq 0, & \forall\, & (i,j) \in \mathcal{C}, \label{eq:gas_compr_flow}\\
& \underline{K}^{\mathrm{C}} \pi_{i,t} \leq \pi_{j,t}\leq \overline{K}^{\mathrm{C}} \pi_{i,t} , & \forall\, & (i,j) \in \mathcal{C}, \label{eq:gas_compr_lim}
\end{align}
\end{subequations}
\vspace{-0.6cm}
\begin{align}
\begin{split}
&\sum_{s \in \Psi^{\mathcal{S}}_i} q^{\mathrm{S}}_{s,t} - \sum_{j \in \Omega_i} m_{ij,t}^{\mathrm{in}} + \sum_{j \in \Omega_i} m_{ji,t}^{\mathrm{out}} \\
& - \sum_{j \in \Omega_i^{\mathrm{C}}} q^{\mathrm{C}}_{ij,t} + \sum_{j \in \Omega_i^{\mathrm{C}}} q^{\mathrm{C}}_{ji,t} - \sum_{j \in \Omega_i^{\mathrm{Cf}}} q^{\mathrm{C}}_{ij,t} \eta^{\mathrm{C}}_c - \sum_{j \in \Omega_i^{\mathrm{Cf}}} q^{\mathrm{C}}_{ji,t} \eta^{\mathrm{C}}_c  \\
& - \sum_{g \in \Psi^{\mathcal{G}}_i}  p^{\mathrm{G}}_{g,t} \eta^{\mathrm{G}}_g - \sum_{d \in \Psi^{\mathcal{D}}_i} (Q^{\mathrm{D}}_{d,t} - q^{\mathrm{O}}_{d,t}) = 0,  \quad 
\forall\,  i \in \mathcal{I}.  \quad  \quad \label{eq:gas_balance}
\end{split}
\end{align}
Constraint \eqref{eq:gas_q_lim_source} restricts gas suppliers to operate between their minimum and maximum, $\underline{Q}_s$ and $\overline{Q}_s$. Constraint \eqref{eq:gas_pres_lim} limits the nodal gas pressure $\pi_{i,t}$ between the minimum $\underline{\Pi}_i$ and maximum $\overline{\Pi}_i$ values. Constraint \eqref{eq:gas_pres_lim_s} fixes the pressures at source nodes, which are directly connected to compressors, denoted by $\mathcal{I}^{\mathrm{S}}$, to a predefined value $\Pi_i^{\mathrm{S}}$.
Constraint \eqref{eq:gas_curtailment} enforces gas load curtailment to be lower than or equal to the nodal gas load $Q^{\mathrm{D}}_{d,t}$. Constraint \eqref{eq:gas_compr_flow} restricts the direction of the flow through compressors, $q^{\mathrm{C}}_{ij,t}$, whereas \eqref{eq:gas_compr_lim} limits the pressure at the outlet of the compressor based on the minimum and maximum compression ratios, i.e., $\underline{K}^{\mathrm{C}}$ and $\overline{K}^{\mathrm{C}}$.
Let $\Omega_i^{\mathrm{C}}$ be the set of compressors starting at node $i$ and $\Omega_i^{\mathrm{Cf}} \subseteq \Omega_i^{\mathrm{C}}$ those that draw fuel from node $i$. Constraint~\eqref{eq:gas_balance} is the nodal gas balance equation, including the gas withdrawal for producing the GFPPs power output $p^{\mathrm{G}}_{g,t}$ with the conversion efficiency $\eta^{\mathrm{G}}_g$, and the gas withdrawal for compressor fuel consumption, calculated as the fraction $\eta^{\mathrm{C}}_c$ of the gas flow through the compressor \cite{WU2000197}. 
\subsection{Gas flow constraints}
The gas flow constraints are enforced for every pipeline $(i,j) \in \mathcal{P}$ and time step $t \in \mathcal{T}$. Binary parameters $U^{\mathrm{QD}}$ and $U^{\mathrm{ST}}$ are used to differentiate between the $\mathrm{DY}$ ($U^{\mathrm{ST}} = U^{\mathrm{QD}} = 1$), $\mathrm{QD}$ ($U^{\mathrm{ST}} = 1, U^{\mathrm{QD}} = 0$) and $\mathrm{ST}$ ($U^{\mathrm{ST}} = U^{\mathrm{QD}} = 0$) models. Equation \eqref{eq:gamma_def_app} is replaced by the feasible set $\mathcal{F}^{\mathbf{S}}$ depending on the chosen solution technique presented in Section~\ref{section:04_NGsol}.

\begingroup
\allowdisplaybreaks
\begin{subequations}\label{eq:OPGF_PDEs}
\begin{align}
    & U^{\mathrm{ST}} \frac{\pi_{ij,t}-\pi_{ij,t-1}}{\Delta t} + \frac{c^2}{A_{ij}}\frac{m_{ij,t}^{\mathrm{out}}-m_{ij,t}^{\mathrm{in}}}{\Delta x}= 0, \label{eq:gas_mass_app} \\ 
        & U^{\mathrm{QD}} \frac{m_{ij,t}-m_{ij,t-1}}{\Delta t} + A_{ij} \frac{\pi_{j,t}-\pi_{i,t}}{\Delta x} + \frac{\lambda_{ij} c^2}{2D_{ij}A_{ij}} \gamma_{ij,t} = 0, \label{eq:gas_momentum_app} \\ 
& m_{ij,t} = \frac{m_{ij,t}^{\mathrm{in}}+m_{ij,t}^{\mathrm{out}}}{2}, \label{eq:gas_m_ave_app} \\ 
& \pi_{ij,t} = \frac{\pi_{i,t}+\pi_{j,t}}{2}, \label{eq:gas_pres_ave_app} \\
& \pi_{ij,0} = \Pi_{ij,0} \label{eq:pressure_initial} \\
& m_{ij,0} = M_{ij,0} \label{eq:flow_initial} \\
& \pi_{ij,0} \leq \pi_{ij,T},  \label{eq:gas_linepack_end} \\
& \underline{M}_{ij} \leq m_{ij,t} \leq \overline{M}_{ij}, \label{eq:m_bound}\\
& \underline{\Gamma}_{ij} \leq \gamma_{ij,t} \leq \overline{\Gamma}_{ij} \label{eq:gamma_bound}.
\end{align}
\end{subequations}
For the first time step, i.e., $t=1$,  \eqref{eq:gas_mass_app} and \eqref{eq:gas_momentum_app} require to assume the initial condition of the average pressure and flow in the pipelines at $t\!-\!1$. To determine the initial condition, we solve model $\mathrm{DY}$-$\mathrm{NLP}$ model for \qty{15}{\minute}
time discretization twice. The first time, we assume that the system starts from a steady-state condition. This implies that, for $t=1$, the steady-state model for \eqref{eq:gas_mass_app} and \eqref{eq:gas_momentum_app} is assumed. We then solve the problem again, assuming the $\mathrm{DY}$ model for all time steps. The solution for the last time step of the first run is used as an initial condition for the second run. After solving the second run, the solution for the last time step is used as an initial condition for problem $\mathrm{P}$-$\mathrm{S}$. Note that we assume that the linepack in each pipeline must be at least restored at the end of the time horizon, i.e., $t = T$,  enforced by \eqref{eq:gas_linepack_end}. Constraints~\eqref{eq:m_bound}--\eqref{eq:gamma_bound} define bounds on the average flow $m_{ij,t}$ and auxiliary variable $\gamma_{ij,t}$, which can be derived as explained in Section~\ref{sec:bounds_mass_flow}.

Finally, the auxiliary variable $\gamma_{ij,t}$ is defined as:
\begin{align}
& \gamma_{ij,t} = \frac{m_{ij,t}|m_{ij,t}|}{\pi_{ij,t}}.\label{eq:gamma_def_app} 
\end{align}

\section{Sequential linear programming}
\label{sec:SLP_app}
Due to the quadratic cost functions of power generators and gas sources, algorithm $\mathbf{P}$-$\mathrm{SLP}$ solves a series of convex quadratic optimization problems. To avoid large step sizes between iterations, trust region or augmentation methods are usually applied to restrict or penalize the step size \cite{ConorThesis}. Here, we apply an augmentation method using the $\ell_2$ norm. We note that in the case of linear cost functions, an $\ell_1$ norm,  as used in \cite{Mhanna2022}, may reduce the total execution time of the algorithm, which we were not able to confirm in our case.

The Euclidean distance between the optimal solution of two consecutive iterations is penalized using a parameter $\delta^k$, which is updated after each iteration. We use $\delta^1 = 10^{-3}$ as an initial value. After each iteration, the penalizer is multiplied by 2, until it reaches a maximum value of $\delta^{\mathrm{max}} = 10^{3}$. The algorithm terminates when the infinity norm of the relative relaxation gap~\eqref{eq:def_relaxation_gap} in iteration $k$, $\Phi^{\infty}_k$ is less than $10^{-6}$. 

\section*{Acknowledgement}
We  express our deepest gratitude to Theis Bo Rasmussen for thoughtful discussions and feedback at the early stages of this research.
 
\bibliographystyle{IEEEtran}
\bibliography{IEEEabrv,ref}

\end{document}